\documentclass{aa}

\usepackage{graphicx}
\usepackage{color}
\usepackage{natbib}
\usepackage{multirow}

\newcommand{\ApJ}[0]{ApJ}

\newcommand{\ApJS}[0]{ApJS}
\newcommand{\AJ}[0]{AJ}
\newcommand{\ARAA}[0]{ARAA}
\newcommand{\AeA}[0]{A\&A}
\newcommand{\MNRAS}[0]{MNRAS}
\newcommand{\AeAS}[0]{A\&AS}

\newcommand{\ASP}[0]{ASP}

\newcommand{\XFe}{X_{\mathrm{Fe}}}
\newcommand{\XFeS}{X_{\mathrm{Fe}\odot}}
\newcommand{\lxtx}{L_{\mathrm X}-T_{\mathrm X}}
\newcommand{\chandra}{{\sl Chandra}}
\newcommand{\cdfs}{{\sl Chandra Deep Field South}}
\newcommand{\correzione}[1]{#1}

\begin{document}

	\title{ICM properties and AGN distribution in high-z RCS clusters}

   \subtitle{}

   \author{Andrea Bignamini\inst{1},
           Paolo Tozzi\inst{2}\inst{,3}, Stefano Borgani\inst{1}\inst{,3}, Stefano Ettori\inst{4} \and
	   Piero Rosati\inst{5}	   
          }

\institute{
$^1$DAUT, Dipartimento di Astronomia dell'Universit\`a di Trieste, via G.B. Tiepolo 11, I--34131, Trieste, Italy \\
$^2$INAF, Osservatorio Astronomico di Trieste, via G.B. Tiepolo 11, I--34131, Trieste, Italy\\
$^3$INFN-- National Institute for Nuclear Physics, Trieste, Italy\\ 
$^4$INAF, Osservatorio Astronomico di Bologna, via Ranzani 1, 40127, Bologna, Italy\\
$^5$ESO, Karl-Schwarzchild Strasse 2, 85748 Garching, Germany\\}

   \date{}

 
\abstract
{The study of the thermodynamical and chemical properties of the Intra
Cluster Medium (ICM) in high redshift clusters of galaxies is a
powerful tool to investigate the formation and evolution of large
scale structures.  Here we discuss the X--ray properties of clusters
of galaxies optically selected in the Red--Sequence Cluster Survey (RCS)
observed with the \chandra\ satellite, at redshifts $0.6 < z < 1.2$.}
{We intended to assess the evolutionary stage of optically selected
high--z clusters of galaxies, performing a spectral analysis of the
diffuse emission from their ICM.  We also investigated the
distribution of AGN in their surroundings.}
{The background subtracted spectra were analyzed and fitted with a
single temperature model to measure average ICM temperature, X--ray
bolometric luminosity and Fe abundance within typical radii between
200 and 350 kpc.  We also analyzed the point source number density and
spatial distribution in the RCS clusters fields as a function of the
X--ray flux.}
{We detected emission for the majority of the clusters, except for
three, for which we have only marginal detection at $\sim3\sigma$.  We
find that the normalization of the Luminosity--Temperature relation for RCS clusters
is a factor of $\sim2$ lower than the one for X--ray selected clusters. We confirm that the
Fe abundance in the detected objects is consistent with that of X--ray
selected clusters at the same redshift.  We also found an excess of
low--luminosity AGN towards the center of the clusters. }
{At $z\sim 1$, optically selected clusters with a well--defined
red--sequence show extended X--ray emission in about 70\% (8/11) of the cases. Nevertheless, their $\lxtx$ relation is significantly lower with respect to X--ray selected clusters at the same redshift, possibly indicating an incomplete virialization. The Fe abundance
measured in the ICM of RCS clusters is comparable to the values
measured for X--ray selected clusters at the same redshift, implying
a substantial enrichment by SNe products.  These two evidences add to
the previously known studies of high--z, X--ray selected clusters, to
confirm that the thermodynamical and chemical properties of the ICM
are substantially established already at $z > 1$. Finally, we find the significant excess of medium and low luminosity AGN close to the centroid of the X--ray emission. Their X--ray emission is not dominating the ICM, but their presence may be relevant for studying the interaction between AGN and ICM.}


\authorrunning{A.\ Bignamini et al.}

\maketitle
%

\section{Introduction}

X--ray selection has been shown to be very effective in finding a very
well--defined sample of clusters of galaxies, with little contamination
and a well--defined selection function.  These properties make X--ray
selection a powerful tool for cosmology \citep[see][]{rbn02}.  On the
other hand, optical selection, based in particular on the presence of
a well--defined red--sequence, is very effective (and much cheaper in
terms of observing strategies) in finding a large number of
candidates.  Recently, this technique has been pushed to high
redshift, as in the case of the Red--Sequence Cluster Survey (RCS), which
uses the color--magnitude relation of early--type galaxies as cluster
finding method \citep{glad00}.

The RCS is a 100 square degree imaging survey using mosaic CCD cameras
on 4m-class telescopes, and is designed specifically to provide a
sample of optically selected $0.2<z<1.2$ clusters. The red--sequence
method is motivated by the observation that all rich clusters have a
population of early--type galaxies which follow a well--defined
color--magnitude relation \citep{yee01}.  The red--sequence represents
the nominal reddest galaxy population in any group of galaxies at the
same redshift, producing a well--defined signature in the
color--magnitude diagram.  So the criterion to find cluster candidates
is based on identifying galaxy over--densities in the four--dimension
space given by color, magnitude and sky coordinate.  For more 
details about the implementation of this method and the survey design
see \citet{glad00,glad01,glad05}, \citet{yee01} and \citet{yee91}.

On the other hand, the X--ray properties of optically selected
clusters are usually different from that of X--ray selected clusters
\citep[see, e.g.][]{D02}.  In particular, lack of X--ray
emission in presence of a well--defined red sequence, can be ascribed
to uncomplete virialization.  In this respect, a systematic study of
the X--ray properties of an optically selected sample of clusters can provide
key information in order to assess both the reliability of an optical
sample for cosmological studies and the dynamical and
thermodynamical status of the ICM in these clusters.

In this Paper we focus on the X--ray properties of clusters of
galaxies in the redshift range $0.6 < z < 1.2$ in the Red--Sequence
Cluster Survey, observed with the \chandra\ satellite.  High
redshift clusters are relevant both for cosmology and for studies of
large scale structure formation and evolution.  First, we investigate
the thermodynamical properties of the ICM, which, given the low number
of photons detected for each object, are described with a
single--temperature model.  Then we also investigate the distribution
of X--ray point sources in the field of RCS clusters. This Paper presents an independent and complementary analysis of X--ray data of the RCS clusters with respect to the Paper of \citet{hick08}. The Paper is
organized as follows.  In \S 2 we describe data reduction and
analysis.  In \S 3 we present our main results on the thermodynamical
and chemical properties of the ICM of RCS clusters.  In \S 4 we
present our results on the AGN distribution in their fields.  Our
conclusions are summarized in \S 5.

\section{Data reduction and analysis} \label{sec:datareduction}

\begin{table*}[htbp]
  \begin{center}
      	\begin{tabular}{|*{7}{c|}}
  	\hline
		Cluster & $z$ & ObsID & Mode & Detector & Exp. [ks] & $N_H$ [10$^{20}$ cm$^{-2}$] \\
\hline
\hline
RCS1419+5326 	& 0.620	& 3240 5886     & VFAINT & ACIS-S & 56.2 & 1.18 \\
\hline
RCS1107.3-0523  & 0.735	& 5825 5887     & VFAINT & ACIS-S & 93.0 & 4.25 \\
\hline
RCS1325+2858	& 0.750	& 3291 4362	& VFAINT & ACIS-S & 61.5 & 1.15 \\
\hline
RCS0224-0002	& 0.778	& 3181 4987	& VFAINT & ACIS-S & 100.9& 2.92 \\
\hline
RCS2318.5+0034  & 0.780 & 4938	        & VFAINT & ACIS-S & 50.0 & 4.14 \\
\hline	
RCS1620+2929	& 0.870	& 3241		& VFAINT & ACIS-S & 33.7 & 2.67 \\
\hline
RCS2319.9+0038  & 0.900 & 5750 7172/3/4 & VFAINT & ACIS-S & 73.7 & 4.19 \\
\hline
RCS0439.6-2905  & 0.960	& 3577 4438     & VFAINT & ACIS-S & 92.0 & 2.64 \\
\hline
RCS1417+5305	& 0.968	& 3239	        & VFAINT & ACIS-I & 62.2 & 1.23 \\
\hline	
RCS2156.7-0448  & 1.080 & 5353 5359     & VFAINT & ACIS-S & 70.7 & 4.60 \\
\hline
RCS2112.3-6326  & 1.099 & 5885	        & VFAINT & ACIS-S & 67.7 & 3.14 \\
\hline
  	\end{tabular}

  \end{center}
  \caption{List of the \chandra\ observations of RCS clusters used
	in this paper.  Redshift are taken from the literature
	\citep{hick08,gilb07}.  The sixth column shows the
	effective exposure times after removal of high background
	intervals.  The last column shows the Galactic $N_H$ values
	measured by \citet{dick90}.}
  \label{tab:rcslist}	
\end{table*}

Data reduction was performed using the CIAO 3.3 software package with
the version 3.2.1 of the Calibration Database (CALDB 3.2.1).  The list
of the \chandra\ Observations of RCS clusters analyzed in this
paper is shown in Table \ref{tab:rcslist}.  The sample consists of all
the public \chandra\ archived observations of RCS clusters to date
(December 2007), for a total of 11 clusters.

We started the data reduction from the level 1 event file.  All the
observations were taken with ACIS--S (except one with ACIS--I) in the
very--faint (VFAINT) mode.  This allowed us to run the tool {\tt
acis\_process\_events} to reduce significantly the instrumental
background, using the values of the pulse heights in the outer 16
pixels of the 5x5 event island.  We also performed a time--dependent
gain adjustment with the tool {\tt acis\_process\_events}.  This
adjustment is necessary because the effective gains of the detectors
are drifting with time as the result of an increasing charge transfer
inefficiency.  The gain--file is used to compute the ENERGY and PI of an
event from the PHA value.  We needed to apply the Charge Transfer
Inefficiency (CTI) correction only for ObsID 3239 taken with ACIS--I.  This
procedure is necessary to recover the original spectral resolution
that is partially lost because of the increasing of CTI due to soft
protons that damaged the ACIS front illuminated chips in the early
phase of the \chandra\ mission.  We filtered the data selecting
events with the standard set of event grades 0, 2, 3, 4, 6.  We also
removed by hand hot columns and flickering pixels.\footnote{We
identify the flickering pixels as the pixels with more than two events
contiguous in time, where a single time interval is set to 3.3 s.}  We
also look for time intervals with high background, by examining the
light--curves of each observation. Usually, we remove only a few
hundreds of seconds of the observing time.  Only in two cases, for
ObsID 3577 and ObsID 5750, we needed to remove a few thousands seconds
from the nominal exposure.

Spectra were extracted from regions corresponding to radii which
maximize the signal--to--noise in the energy range 0.5--6.0 keV.  The extraction radii are in the range between
25 and 45 arcsec or between 200 and 350 kpc. After a visual
inspection, we removed the contribution of point sources, most of them
low--luminosity AGN, embedded in the ICM diffuse emission (not
necessarily related to the cluster) as suggested by \citet{bran07b}.
For details about the point sources identification and their flux
contribution see \S 4.

Background subtraction must be done accurately, given the low surface
brightness of the sources.  However, this procedure is simplified by
the fact that the emission of each source is always within a radius of 40
arcsec.  Therefore, we could sample the background from the same
observation, in a region typically three times larger by size than the
extraction region of the source spectra.  Indeed, in our case a
different procedure involving the creation of synthetic background
spectra to match exactly the same position of the source on the
detector, would be risky, especially due to fluctuations in the soft
band associated with the Galactic emission.  Calibration files (rmf and
arf) were built for the extraction regions, while soft and hard
monochromatic exposure maps (for energies of 1.5 and 4.5 keV
respectively) are used to compute aperture photometry of point sources
(see \S 4).

The background subtracted spectra were analyzed with {\it Xspec}
v.12.3.0 \citep{arna96}.  Since the signal--to--noise ratio for our
clusters is low, we use the {\sl C-statistic} as criterion to find the
best fit models \citep{bevi02,arna04}.  In our spectral fits there
are three free parameters: temperature, metallicity and normalization.
We fitted the spectra with a single temperature {\tt mekal} model
\citep{kaas92,lied95} and model the Galactic absorption with {\tt
tbabs} \citep{wilm00}, fixing the Galactic neutral Hydrogen column
density ($N_H$ in Table \ref{tab:rcslist}) to the value
obtained with radio data \citep{dick90}.  Redshifts are fixed to the
values measured from the optical spectroscopy \citep{hick08,gilb07}.

It has recently been shown that a methylene layer on the \chandra\
mirrors increases the effective area at energies larger than 2 keV
\citep[see][]{mars03}
This has a small effect on the total measured fluxes, but it can have
a non--negligible effect on the spectral parameters.  To correct for
this, we included in the fitting model a ``positive absorption edge''
({\it Xspec} model {\tt edge}) at an energy of 2.07 keV and with $\tau
= -0.17$ \citep{vikh05}.  This multiplicative component artificially
increases the hard fluxes by $\simeq 3.5$\%, therefore the final hard
fluxes and luminosities computed from the fit are corrected downwards
by the same amount.

The ratio between the elements are fixed to the solar values as in
\citet{ande89}. These values for solar metallicity have been
superseded by the new values of \citet{grev98} and \citet{aspl05}, who
found a 0.676 and 0.60 times lower Iron solar abundance
respectively. However, we prefer to report Fe abundances in units of
solar abundances by \citet{ande89}, because most of the literature
still refers to them. Since our measures of metallicity are not
affected by the presence of other metals, because the only detectable
emission lines are the H--like and He--like Iron $K_\alpha$ line
complex at rest--frame energies of 6.6--6.9 keV, the updated values
can be obtained simply rescaling the values reported in Table
\ref{tab:rcsresults} by the factor 1/0.676 or 1/0.60.

The fits were performed over the energy range 0.5--6.0 keV.
Net counts are extracted in the same energy range- We removed
low energy photons in order to avoid uncertainties in the ACIS
calibration at low energies. The cut at high energies, instead, is
imposed by the rapidly decreasing signal--to--noise ratio at energies
larger than 6.0 keV, due to the combination of the lower effective area
of ACIS, and of the exponential cut--off of the high--z, thermal
spectra.

Finally, we computed the X--ray bolometric luminosities with {\it Xspec}
integrating over the entire X-ray band the analytical function
describing the best fit of each spectrum, and adopting a
$\Lambda$CDM cosmology with $\Omega_m$ = 0.3, $\Omega_{\Lambda}$
= 0.7 and $H_0$ = 70 km s$^{-1}$ Mpc$^{-1}$.

\section{Results on the Properties of the ICM}

In this Section we present and discuss the results of the X--ray
spectral analysis of the high--z RCS clusters.  Spectra are extracted
within a radius chosen in order to maximize the signal--to--noise
ratio in the 0.5--6.0 keV band.  The extraction radii and the number of
net counts detected for each cluster are shown in Table
\ref{tab:rcsresults}.  These values are obtained with simple aperture
photometry, by subtracting the total number of events in the background
extraction region scaled by the area ratio from the number of events
in the cluster region.  The net counts error is obtained from the
Poissonian error of the numbers of counts.

Best fit temperatures, Fe abundances, fluxes and bolometric luminosities are
also shown in Table \ref{tab:rcsresults}; error bars refer to 1
$\sigma$ confidence levels.  We show that we are able to measure
temperatures with a typical 1 $\sigma$ error bar of 20-30\%.  Errors
on luminosities are taken from the Poissonian error on the net detected
counts.

For three clusters (RCS1417, RCS2112 and RCS2156), we have only
marginal detection of the diffuse emission consistent with noise
within $\sim3\sigma$.  So in these three cases we do not compute the
maximum signal--to--noise region, but simply selected a circular
region with radius of $\sim20$ arcsec centered in the optical
coordinates of the clusters.  Since we are not able to perform
spectral analysis for these clusters, we only provide an upper limit
for the bolometric luminosities corresponding to a temperature range
of 1.0--8.0 keV and a fixed metallicity values ($0.3 \XFeS$).

In order to obtain a more robust measure of the diffuse emission for
these three clusters, we merged together the three X--ray images
overlapping the optical centers of the clusters.  From the merged
image (Figure \ref{fig:invisible}) we found \hbox{$274\pm 29$} total
net counts and a mean bolometric luminosity for each cluster
$L_{\mathrm X}\sim0.2\times10^{44}$ erg s$^{-1}$. This implies that on
average the three clusters show extended emission.  We
are not able to derive an average temperature from the combined
spectrum, due to the low number of net counts.

As one can see in the Table \ref{tab:rcslist} and Table
\ref{tab:rcsresults} the cluster RCS1417 is the only one observed
with ACIS--I and the one with the lowest number of net counts.
Despite this it is not the one with the lowest signal--to--noise
ratio. This is mostly due to the lower background of the front
illuminated chips of ACIS--I with respect to that of the back illuminated chip
of ACIS--S.  Still, RCS1417 is formally undetected in the ACIS--I
image.

The detailed spectral analysis of each cluster along with the X--ray
spectra are shown in Appendix  A.

\subsection{The Luminosity--Temperature Relation}

\begin{figure}
\centering
\includegraphics[width=8.0 cm, angle=0]{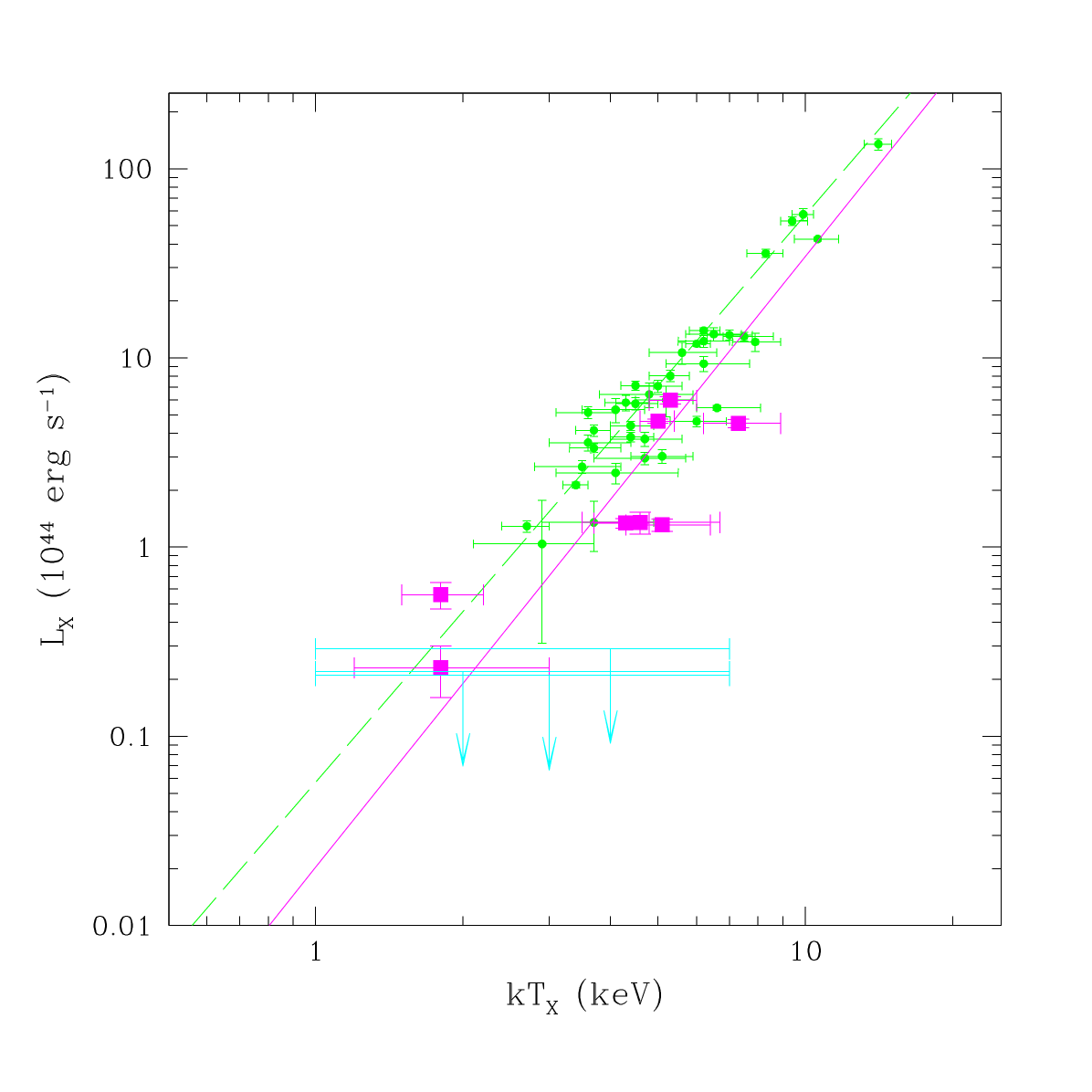}
\caption{$\lxtx$ relation for the RCS sample
(magenta squares; upper limits to X--ray luminosity are shown in cyan
for RCS clusters without X-ray detection). The solid magenta line is the best fit of the $\lxtx$ for our RCS sample.
The green points are the X--ray selected sample data by \citet{bran07c} and the dashed green line is their best fit.}
\label{fig:lt}
\end{figure}

\begin{table*}[htpb]
\begin{center}
\tiny{  	\begin{tabular}{|*{10}{c|}}
  	\hline
\multirow{2}{10ex}{\centering{Cluster}} & $r_{ext}$ & $r_{ext}$ & \multirow{2}{10ex}{\centering{Net counts}} & \multirow{2}{5ex}{\centering{SNR}} & $kT$ & \multirow{2}{12ex}{\centering{$X_{\mathrm{Fe}}/X_{\mathrm{Fe}\odot}$}} & $S_{0.5-2.0}$ & $S_{2.0-7.0}$ & $L_{\mathrm X}$\\
& [arcsec] & [kpc] & & & [keV] & & \tiny{[$10^{-14}$erg/cm$^{2}$/s]} & {[10$^{-14}$erg/cm$^{2}$/s]} & [$10^{44}$erg/s]\\
\hline
\hline
RCS1419+5326   & 37.05 & 252 & $2320\pm60$ & 38.0 & $ 5.0_{-0.4}^{+0.4} $ & $0.29_{-0.11}^{+0.06}$ & $10.9\pm0.3$ & $11.6\pm0.6$ & $4.63\pm0.12$ \\
\hline			  
RCS1107.3-0523 & 28.43 & 207 & $710\pm40$  & 15.5 & $ 4.3_{-0.6}^{+0.5} $ & $0.67_{-0.27}^{+0.35}$ & $2.26\pm0.12$ & $2.0\pm0.2$ & $1.34\pm0.08$ \\
\hline			  
RCS1325+2858   & 29.73 & 218 & $90\pm30$   & 2.8  & $ 1.8_{-0.6}^{+1.2} $ & $0.09_{-0.09}^{+0.66}$ & $0.43\pm0.08$ & $0.09\pm0.07$ & $0.23\pm0.07$ \\
\hline			  
RCS0224-0002   & 36.69 & 273 & $740\pm50$  & 13.1 & $ 5.1_{-0.8}^{+1.3} $ & $0.01_{-0.01}^{+0.14}$ & $1.85\pm0.12$ & $1.7\pm0.2$ & $1.31\pm0.10$ \\
\hline			  
RCS2318.5+0034 & 40.56 & 302 & $970\pm50$  & 19.1 & $ 7.3_{-1.0}^{+1.3} $ & $0.35_{-0.22}^{+0.20}$ & $5.4\pm0.3$ & $7.5\pm0.6$ & $4.51\pm0.23$ \\
\hline			  
RCS1620+2929   & 29.45 & 227 & $190\pm20$  & 7.5  & $ 4.6_{-1.1}^{+2.1} $ & $0.33_{-0.33}^{+0.60}$ & $1.51\pm0.17$ & $1.3\pm0.3$ & $1.35\pm0.18$ \\
\hline			  
RCS2319.9+0038 & 45.62 & 356 & $1490\pm60$ & 22.5 & $ 5.3_{-0.5}^{+0.7} $ & $0.60_{-0.18}^{+0.22}$ & $5.8\pm0.2$ & $5.9\pm0.4$ & $5.97\pm0.26$ \\
\hline			  
RCS0439.6-2905 & 24.71 & 196 & $220\pm30$  & 5.7  & $ 1.8_{-0.3}^{+0.4} $ & $0.44_{-0.27}^{+0.27}$ & $0.65\pm0.06$ & $0.11\pm0.07$ & $0.56\pm0.09$ \\
\hline			  
RCS1417+5305   & 19.68 & 156 & $37\pm11$   & 2.9  & 1.0--8.0              & (0.3)                  & $<0.21$ & $<0.16$ & $<0.29$       \\
\hline			  
RCS2156.7-0448 & 19.68 & 160 & $60\pm20$   & 2.4  & 1.0--8.0              & (0.3)                  & $<0.10$ & $<0.07$ & $<0.22$       \\
\hline
RCS2112.3-6326 & 19.68 & 161 & $47\pm20$   & 2.0  & 1.0--8.0              & (0.3)                  & $<0.12$ & $<0.19$ & $<0.21$       \\
\hline				     
\end{tabular}
}
  \end{center}	
	\caption{Best fit values of the spectra of each cluster.
	  Errors are at 1$\sigma$ confidence level. For each cluster is shown: extraction radius in arcsec and kpc; net counts in the 0.5--6.0 keV band; signal--to--noise ratio; best fit temperature; best fit Iron abundance; fluxes in the soft (0.5--2.0 keV) and hard (2.0--7.0 keV) band; best fit bolometric luminosity. Bolometric luminosities are computed for $\Omega_m$ = 0.3, $\Omega_{\Lambda}$ = 0.7 and $H_0$ = 70 km s$^{-1}$ Mpc$^{-1}$.}
	\label{tab:rcsresults}
\end{table*}

\begin{figure}
\centering
\includegraphics[width=8.0 cm, angle=0]{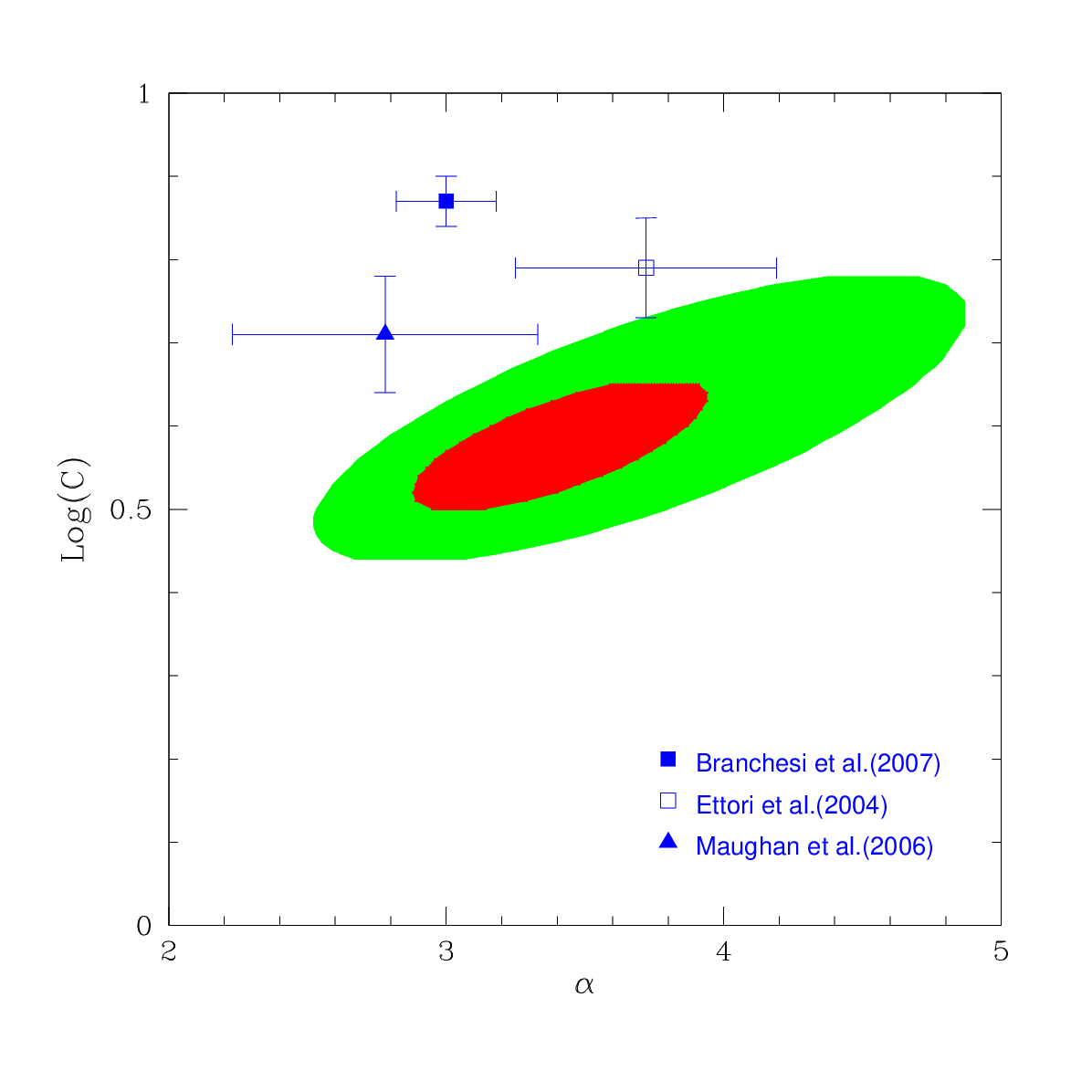}
\caption{Comparison between the RCS sample and X--ray selected samples $\lxtx$ best fit. The red (green) region is 1$\sigma$ (2$\sigma$) confidence level for RCS clusters. Blue filled square is the best fit for the sample by \citet{bran07c}, empty square for \citet{etto04} and filled triangle for \citet{maug06}. For all these fits the luminosities are scaled by the cosmological factor $E(z)^{-1}(\Delta_c(z)/\Delta_c(z=0))^{-1/2}$.}
\label{fig:contourfit}
\end{figure}

\begin{figure}
\centering
\includegraphics[width=8.0 cm, angle=0]{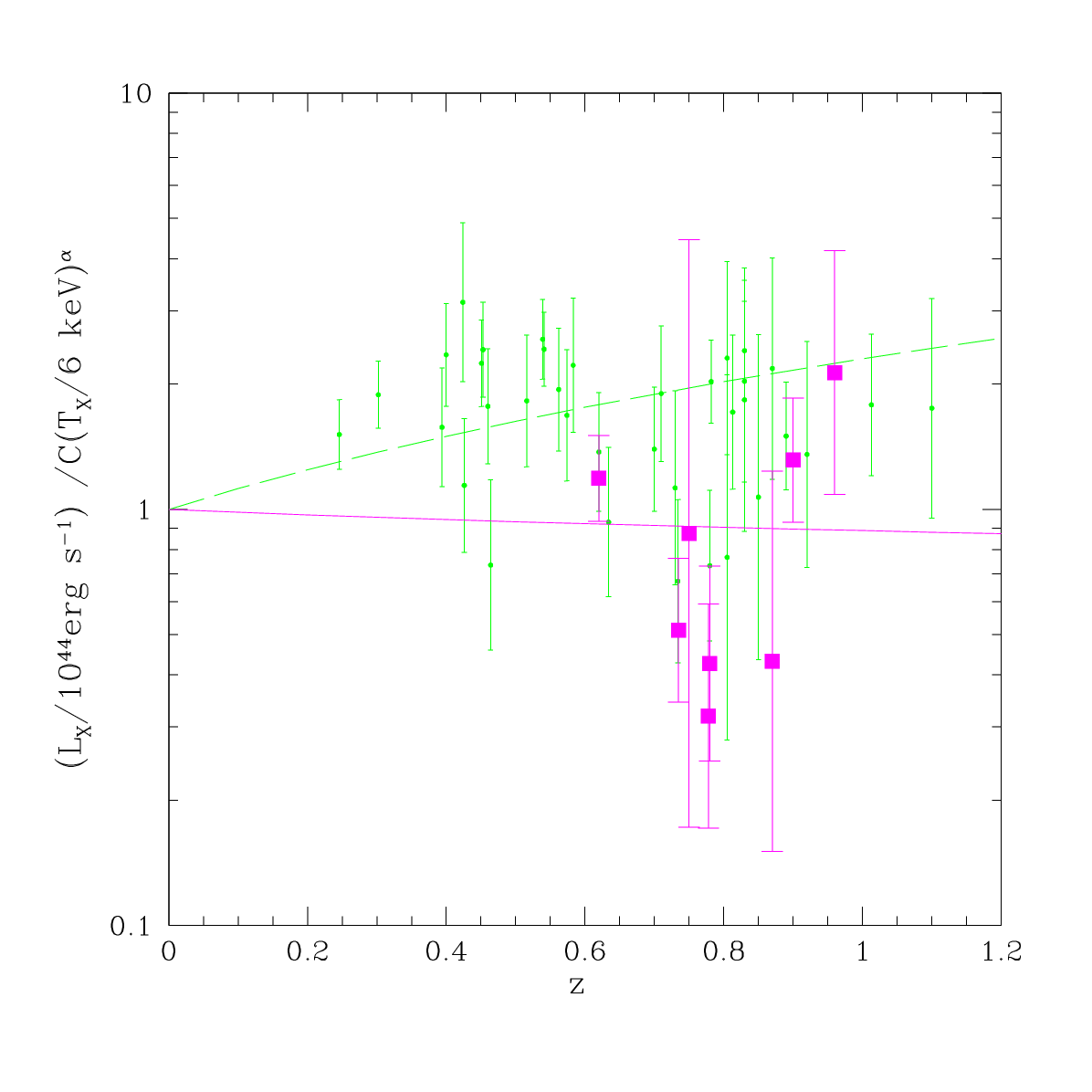}
\caption{Evolution of the $\lxtx$. Ratio of the observed luminosity to the expected luminosity form the local relation as in \citet{mark98} versus redshift. The slope and the normalization of the $\lxtx$ relation for local clusters
are fixed to be $\alpha=2.64$ and $logC = 0.80$ respectively. Magenta squares are RCS data and the solid magenta line is the best fit. Small green points are data from \citet{bran07c} and the dashed green line is their best fit.}
\label{fig:lsut}
\end{figure}

We checked whether the RCS clusters follow the same $\lxtx$
 relation observed for X--ray selected clusters. In Figure
\ref{fig:lt} we compare the bolometric luminosity and the temperature measured
for RCS clusters to the best fit of the $L_{\mathrm
X}-T_{\mathrm X}$ relation as obtained by \citet{bran07c}.

\textcolor{black}{We compared the slope and normalization of the $\lxtx$ relation of RCS
clusters with that of X--ray selected high--z clusters.
We fitted the $\lxtx$ only for the 8 detected clusters with a single power law assuming the following expression:
\begin{equation}
L_{44}=CT_6^\alpha,
\end{equation}
where $L_{44}$ is the bolometric luminosity in units of \hbox{$10^{44}$ erg s$^{-1}$} and $T_6=kT/$(6 keV).
Our best fit is $Log C = 0.82^{+0.08}_{-0.08}$ and $\alpha=3.2^{+0.7}_{-0.4}$, while best fit of \citet{bran07c} is $Log C = 1.06 \pm 0.03$ and $\alpha=3.00^{+0.19}_{-0.18}$. We found that the slope is in agreement within the fairly large uncertainties between the two sample, whereas the RCS normalization is about a factor of $\sim2$ lower at high temperature and about a factor of $\sim3$ lower at low temperature. The two fits are inconsistent with a significance of $\sim4.2\sigma$.}

The three marginally detected RCS clusters, for which we
plot in Figure \ref{fig:lt} only the upper limits on the luminosity for a wide temperature
range 1.0--8.0 keV, seem to have lower luminosities with respect to the
fitted RCS $L_{\mathrm X}-T_{\mathrm X}$, but only if they have $kT>3$ keV.
Unfortunately, we are not able to estimate their average temperature
given the low number of net detected counts.

\citet{bran07c} repeated the analysis after removing the expected self--similar evolution in the $\lxtx$ relation. In order to compare our sample with this analysis and with other authors (namely \citealt{etto04} and \citealt{maug06}), we removed
the expected self--similar evolution from the luminosities of our
cluster sample, i.e.\ scaling the luminosity by the cosmological factor
$E(z)^{-1}(\Delta_c(z)/\Delta_c(z=0))^{-1/2}$. The evolution of the Hubble parameter with redshift reads
\begin{equation}
E(z)^2=\Omega_m(1+z)^3+\Omega_\Lambda
\end{equation}
and $\Delta_c(z)$ is the average density of virialized objects in units of the critical density \citep{eke96}.
A simple, but accurate expression for $\Delta_c(z)$ is given by \citet{brya98}.

Again, we fitted the $\lxtx$ only for the 8 detected clusters with a single power law, now assuming the following expression:
\begin{equation}
 E(z)^{-1}(\Delta_c(z)/\Delta_c(z=0))^{-1/2}L_{44}=CT_6^\alpha.
\end{equation}

The parameters $LogC$ and $\alpha$, determined by the best fit, are listed in Table \ref{tab:bestfits}
compared with those of \citet{bran07c}, \citet{etto04} and \citet{maug06}.

\begin{table}[!h]
\centering
\begin{tabular}{|c|c|c|}
\hline
Sample                & $\alpha$      & $LogC$\\
\hline
\hline
RCS                   & $3.3_{-0.4}^{+0.6}$    & $0.57\pm0.08$\\
\hline
\citeauthor{bran07c}  & $3.00_{-0.18}^{+0.19}$  & $0.87\pm0.03$\\
\hline
\citeauthor{etto04}   & $3.72\pm0.47$           & $0.79\pm0.06$\\
\hline
\citeauthor{maug06}   & $2.78\pm0.55$           & $0.71\pm0.07$\\
\hline
\end{tabular} 
\caption{Slope and normalization of the $\lxtx$ relation for RCS and other X--ray selected clusters after removing the expected self--similar evolution.}
\label{tab:bestfits}
\end{table}

\textcolor{black}{Also with luminosities scaled by the expected self--similar evolution, we find that the slope is in good agreement with other authors, while always
RCS clusters show a factor of $\sim2$ lower normalization of the $\lxtx$ relation
at high confidence level (samples inconsistent at $4.0\sigma$ for \citealt{bran07c}, $2.6\sigma$ for \citealt{etto04} and $1.9\sigma$ for \citealt{maug06}) with respect to that of X--ray selected clusters at similar redshift. Figure \ref{fig:contourfit} summarizes this comparison, showing the $1\sigma$ and the $2\sigma$ confidence level for our fit against the best fit values with corresponding $1\sigma$ error for the X--ray selected samples}.
The independent analysis by \citet{hick08} found as well that RCS clusters are about a factor of two less luminous for a given temperature than X--ray selected clusters, however their best fit slope is significantly flatter than ours.

\correzione{\citet{lubi04} presented a detailed analysis of two high--z optically selected clusters. The X--ray properties of both clusters are consistent with the high--redshift $\lxtx$ relation measured from X--ray selected samples. However, based on the local relations, their X--ray luminosities and temperatures are low, by a factor of 2--9, for their measured velocity dispersion. The exact cause of these results is unclear. The authors claimed that the differences in X--ray properties of these two clusters may result from the fact that these clusters at these epochs are still in process of forming.}

The lower luminosity at fixed temperature of RCS clusters with respect to X--ray selected clusters, can not be entirely ascribed to our different choice of the extraction radius, which results in bolometric luminosity estimated within smaller radius (see below for details about extraction radii and bolometric luminosity measures adopted by other authors). Actually, adopting the $\beta$ model fit by \citet{hick08} for RCS clusters, we estimate on average only $\sim20\%$ the luminosity loss due to our smaller radius choice.

We also analyzed the evolution with redshift of the $\lxtx$
relation for our sample of RCS clusters.
In Figure \ref{fig:lsut}, we show the quantity $L_{\mathrm X}/CT_{\mathrm X}^{\alpha}$, where $\alpha$ and $C$ are fixed to values for the local $\lxtx$ relation as in \citet{mark98}, $\alpha = 2.64$ and  $LogC=0.80$.
We fitted the evolution of the observed
$L_{\mathrm X}/CT_{\mathrm X}^{\alpha}$ as a function of redshift with
a single power law, assuming the following expression:

\begin{equation}
L_{44}/CT_6^\alpha=(1+z)^A.
\end{equation}

From the best fit we found $A=-0.2\pm0.2$.  The slope of the best fit is well
consistent with zero, implying no evolution in the $\lxtx$.

We note that, as pointed out by \citet{sant08}, only the lowest
redshift cluster RCS1419 shows a cool--core. \textcolor{black}{Therefore we repeated the spectral analysis masking the cool--core in
RCS1419. In this case the temperature obtained by the best fit is
$kT=5.2^{+0.7}_{-0.5}$ keV. We corrected the bolometric luminosity
for the removed core emission fitting the radial surface brightness profile with a $\beta$ model
and extrapolating the profile in the masked inner region.
The corrected bolometric luminosity is $L_{\mathrm
X}=(3.71\pm0.15)\times10^{44}$ erg s$^{-1}$.}
Finally, we repeated the fit using
these values for RCS1419 and we found $L_{\mathrm X}/T_{\mathrm X}^{\alpha} \propto
(1+z)^{-0.4 \pm 0.2}$. 
In this case we have evidence of slightly negative evolution for the $\lxtx$ for RCS clusters, consistent with no evolution at $2\sigma$ confidence level.

To understand properly the meaning of this result, a rigorous comparison with various studies about X--ray selected clusters can be useful, since the evolution of the $\lxtx$ relation for X--ray selected clusters was frequently explored in last years, in some cases deriving conclusions slightly varying with authors.

\citet{vikh02} analysed 22 clusters observed with \chandra\ at redshift between 0.4 and 1.26. Temperatures were measured by fitting a spectrum integrated within a radius of $0.35-0.70 h^{-1}_{70}$ Mpc with a single temperature mekal model. Bolometric luminosities were extrapolated with a $\beta$ model to a fixed radius of 1.4 $h^{-1}_{70}$ Mpc, excluding the central 70$h^{-1}_{70}$kpc region for clusters with sharply peaked surface brightness profiles. \citet{vikh02} claimed a positive evolution, with $A=1.5\pm0.2$.

A similar result was obtained by \citet{lumb04} and \citet{koto05} who analysed respectively 10 and 8 XMM observed clusters in a smaller and lower redshift range, $0.4<z<0.7$. \citet{lumb04} adopted a mekal model to extrapolate luminosities within a virial radius, $r_v$, according to the $T-r_v$ relation of \citet{evra96}. For each cluster best fit temperatures from spectral fits within $120''$ were used. \citet{lumb04} found a positive evolution of the $\lxtx$, $A=1.52^{+0.26}_{-0.27}$. \citet{koto05}, instead, evaluated bolometric luminosities within r$<$ 1400 kpc, using a mekal model with emission--weighted temperature and correcting the inner 70 kpc emission with a best fit $\alpha$--$\beta$ model. Their $\lxtx$ shows a positive evolution $A=1.8\pm0.3$.

\citet{maug06} analysed the evolution with redshift of the $\lxtx$, using 11 clusters observed with \chandra\ or XMM at redshift $0.6<z<1.0$. They adopted $L_{200}$ luminosities obtained in the following way. First the surface brightness distribution of each cluster was modeled with a two--dimensional $\beta$ model within a detection radius; then the luminosities were extrapolated to $R_{200}$, according to the best fitting surface brightness profiles. Comparing with the local $\lxtx$ of \citet{mark98}, \citet{maug06} found a positive evolution $A=0.7\pm0.4$

A sample 28 clusters observed with \chandra\ at redshift between 0.4 and 1.3 was studied by \citet{etto04}, who extrapolated bolometric luminosities within $R_{500}$, according to a $\beta$ model fitted within a radius which optimize the signal--to--noise ratio. When they use their whole sample they found a positive evolution $A\sim0.6$ for the $\lxtx$, instead when they use only clusters with redshift greater than 0.6 (16 objects) they found a lower positive evolution $A\sim0.1$. The authors explored widely the causes of this relevant differences with other authors, in particular with respect to the similar redshift range sample by \citet{vikh02}. Basically, beyond differences in fitting procedure, definition of reference radii and temperature estimation, the sample of \citet{etto04} is larger at higher redshift, 16 galaxy clusters at $z>0.6$ and 4 at $z>1.0$, whereas \citet{vikh02} has 9 and 2 objects respectively, suggesting a different evolution of the $\lxtx$ relation according to different redshift range studied.

A similar conclusion was also derived by \citet{bran07c}, who used a sample of 17 clusters from che \chandra\ archive supplemented with additional clusters from \citet{vikh02}, \citet{etto04} and \citet{maug06}, to form a final sample of 39 high redshift, $0.25<z<1.3$, clusters. Their luminosities were extrapolated to $R_{500}$ using an isothermal $\beta$--profile. For their whole combined sample, \citet{bran07c} found a positive evolution $A=1.20\pm0.08$. However, the $\chi^2$ of this fit has a probability lower than 0.1\% to be acceptable, suggesting that the evolution of the $\lxtx$ on the whole redshift range cannot be described by any power low of the form $\propto (1+z)^A$. In particular, they claimed that a stronger evolution is required at lower redshift, namely up to $z\sim0.6$, followed by a much weaker evolution at higher redshift, as it is shown by their data over--plotted in Figure \ref{fig:lsut}.

\begin{figure}
\centering
\includegraphics[width=8.0 cm, angle=0]{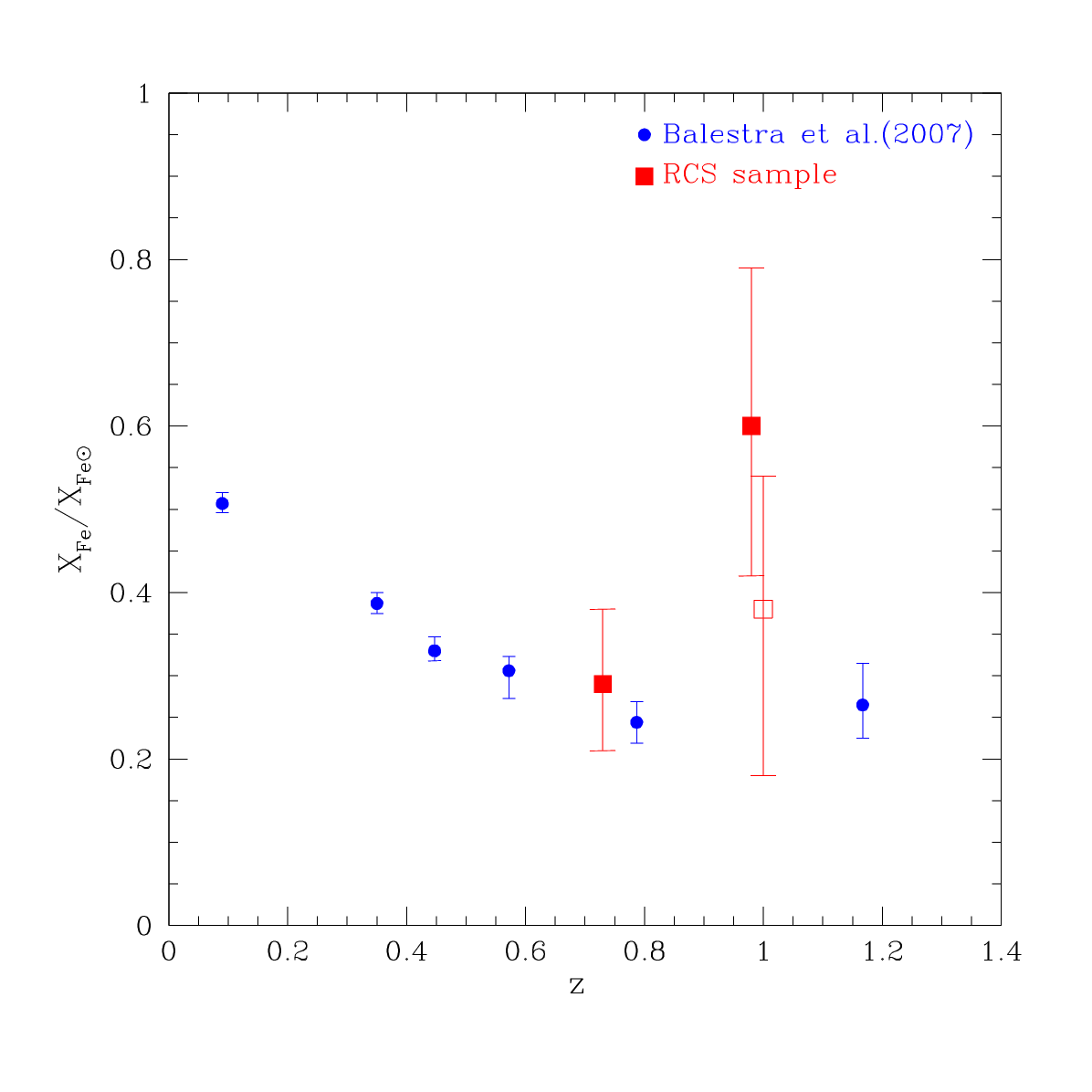}
\caption{Mean Iron abundance of RCS sample (red squares) in two redshift
bins compared with the results of \citet{bale07} (blue circles).
The open square represent the second bin without RCS2319.}
\label{fig:metvsredshift}
\end{figure}

\begin{table*}[!ht]
  \small{
    \begin{center}
      \begin{tabular}{|*{7}{c|}}
\hline
        &  \multirow{3}{14ex}{\centering Solid angle [deg$^2$]}       & \multicolumn{3}{|c|}{Number} & \multirow{3}{18.5ex}{\centering Flux limit (soft) [erg/cm$^2$/s]} & \multirow{3}{18.5ex}{\centering Flux limit (hard) [erg/cm$^2$/s]} \\
Cluster	&         & \multicolumn{3}{|c|}{ of sources  } &                                                                   &                                                                   \\
\cline{3-5}
        &         & soft & hard  & tot            &                                                                   &                                                                   \\			
\hline
\hline
RCS1419+5326   & $0.025$ & 31  & 19 & 43 & $3.4\times 10^{-16}$ & $2.8\times 10^{-15}$ \\
\hline
RCS1107.3-0523 & $0.025$ & 48  & 33 & 56 & $2.3\times 10^{-16}$ & $1.9\times 10^{-15}$ \\
\hline
RCS1325+2858   & $0.025$ & 35  & 23 & 40 & $3.2\times 10^{-16}$ & $2.6\times 10^{-15}$ \\
\hline
RCS0224-0002   & $0.024$ & 47  & 35 & 70 & $2.1\times 10^{-16}$ & $1.7\times 10^{-15}$ \\
\hline
RCS2318.5+0034 & $0.021$ & 29  & 17 & 40 & $3.6\times 10^{-16}$ & $2.7\times 10^{-15}$ \\
\hline
RCS1620+2929   & $0.021$ & 25  & 18 & 34 & $2.5\times 10^{-16}$ & $2.0\times 10^{-15}$ \\
\hline	
RCS2319.9+0038 & $0.023$ & 38  & 26 & 43 & $2.5\times 10^{-16}$ & $1.9\times 10^{-15}$ \\
\hline
RCS0439.6-2905 & $0.024$ & 30  & 26 & 45 & $2.5\times 10^{-16}$ & $2.1\times 10^{-15}$ \\
\hline	
RCS1417+5305   & $0.082$ & 114 & 88 & 167 & $1.0\times 10^{-16}$ & $1.0\times 10^{-15}$ \\
\hline
RCS2156.7-0448 & $0.024$ & 34  & 18 & 43 & $2.6\times 10^{-16}$ & $2.2\times 10^{-15}$ \\
\hline
RCS2112.3-6326 & $0.021$ & 38  & 26 & 48 & $3.2\times 10^{-16}$ & $2.4\times 10^{-15}$ \\
\hline
\end{tabular}

    \end{center}}	
	\caption{Solid angle and number of identified point sources in the soft (0.5-2.0 keV) and
	hard (2.0-7.0 keV) band, with the corresponding flux limits in each field.}
	\label{tab:rcssources}
\end{table*}

From this concise summary of studies based on high-z X--ray selected clusters, it is evident that the evolution of the $\lxtx$ relation is still topic of debate both for significant differences in measuring luminosities and/or temperature and in result interpretations.

Besides, our claim of no evolution or slightly negative evolution for the $\lxtx$ for RCS clusters is bounded by this final remark:
we can only compare our $\lxtx$ for high--z optically selected clusters with local X--ray based $\lxtx$, whereas it would be better to compare our clusters with similarly selected local ones, in order to understand the physical properties evolution of optically selected clusters.

However, since the evolution is systematically and significantly lower, even this analysis can be a further indication of lower normalization of the $\lxtx$ for high--z optically selected clusters.

It is clear that in order to asses the evolution of the $\lxtx$ relation for optically
selected clusters, we need to collect X--ray data for a larger number
of objects distributed on a wider redshift range.

\subsection{The Iron Abundance}

As one can see in Table \ref{tab:rcsresults}, in the majority of cases
the Fe abundances from the spectral analysis of single clusters are
consistent with zero within 1$\sigma$ and all, except those of RCS1620
and RCS2319, are consistent with zero within 2$\sigma$.  This is
mostly due to the low signal--to--noise ratio of the spectra.  To
obtain a more robust measurement of the average Fe abundance, we
fitted at the same time all the spectra using a single value for the
Iron abundance. In this way we could increase the signal--to--noise
ratio and measure the mean Fe abundance of the sample.  This has been
already done to measure the evolution of the average Fe abundance at
high--z \citep[see][]{tozz03,bale07}. Like in the single cluster
analysis, we fix the local absorption to the Galactic neutral
Hydrogen column density and the redshift to the value measured from the
optical spectroscopy.  We also fix the temperatures to the values
obtained previously from the single cluster analysis, except for
RCS1417, RCS2112 and RCS2156, whose temperatures vary in the range
1.0--8.0 keV.  The combined fit analysis allowed us to measure the average
Fe abundance which turns out to be $\langle \XFe \rangle
=0.37_{-0.08}^{+0.09}\XFeS$.  Therefore we detect with high
significance the presence of Iron at a level consistent with that of
X--ray selected clusters at similar redshift.

We also repeated the same procedure after dividing our sample into two
redshift bins. The first bin includes all clusters with redshift lower
than 0.80 including 5 object with about 4800 total net counts and $<$$z$$>\
\simeq 0.73$, whereas the second bin includes 6 objects with redshift
$z>0.8$ with about 2000 total net counts and $<$$z$$>\ \simeq0.98$.
The spectral information in this bin is largely dominated by RCS2319,
which shows an Iron abundance slightly larger than the typical value at
$z\sim 1$.  From the combined fit we found $\langle \XFe \rangle
=0.29_{-0.08}^{+0.09}\XFeS$ for the first subsample and $\langle \XFe
\rangle =0.60_{-0.18}^{+0.19}\XFeS$ for the second.  Excluding RCS2319
from the second bin, we found $\langle \XFe \rangle =
0.43_{-0.20}^{+0.16}\XFeS$.  These values are plotted in Figure
\ref{fig:metvsredshift} and, given the large error bars, are in agreement with the results of \citet{bale07}.  It is clear that, in
order to investigate the typical Fe abundance in the ICM of
RCS clusters at high redshift, we need to use a
substantially larger sample, observed with medium deep \chandra\
exposures.  However, our results show that also for this
sample of optically selected clusters, the ICM was already enriched
with Iron at a level comparable with that of X--ray selected clusters.

\section{Point Sources} \label{sec:ps}

\begin{table*}[htpb]
  \small{
    \begin{center}
      \begin{tabular}{|*{5}{c|}}
\hline
\multirow{2}{3ex}{\centering {\#}}   & \multirow{2}{10ex}{\centering {Cluster}} &  CL hardness & PS hardness & $L_{\mathrm X}$ \\
                           &   & ratio & ratio & [$10^{44}$ erg/s] \\
\hline
\hline
 1 & RCS1419+5326    &   0.0 &  0.1 & 0.60 \\
\hline
 2 & RCS1107.3-0523  &  -0.1 & -0.2 & 0.15 \\
\hline
 3 & RCS1107.3-0523  &  -0.1 &  0.1 & 0.19 \\
\hline
 4 & RCS1325+2858    &  -0.7 &  0.9 & 0.09 \\
\hline
 5 & RCS0224-0002    &   0.0 &  1.0 & 0.05 \\
\hline
 6 & RCS2318.5+0034  &   0.1 &  0.6 & 0.21 \\
\hline	
 7 & RCS2318.5+0034  &   0.1 &  0.3 & 0.09 \\
\hline	
 8 & RCS1620+2929    &  -0.1 &  0.2 & 1.08 \\
\hline
 9 & RCS2319.9+0038  &   0.0 &  0.3 & 0.29 \\
\hline
10 & RCS0439.6-2905  &  -0.7 &  0.4 & 0.17 \\
\hline
11 & RCS1417+5305    &  -0.2 &  1.0 & 0.14 \\
\hline	
12 & RCS2156.7-0448  &  -0.2 &  1.0 & 0.31 \\
\hline

\end{tabular}
    \end{center}}	
	\caption{Comparison between the hardness ratio of the clusters (CL)
	and the hardness ratio of the point sources (PS) identified within
	150 kpc from the corresponding cluster center.  The last
	column shows the luminosity in the 0.5--10.0 keV band for
	point sources assumed to be at the same redshift of the
	cluster.}
	\label{tab:rcssources2}
\end{table*}

To study the large scale structure associated with clusters we
computed the number density of active galactic nuclei (AGN) in the
regions around RCS clusters, to be compared with the AGN density in
the field.  An interesting aspect, currently under investigation, is
whether the nuclear activity is enhanced in galaxies that belong to
high density structures, like clusters or filaments.  These effects are
difficult to investigate, and preliminary works on deep X--ray fields
offered only a tantalizing hint for an enhanced AGN activity
\citep{gill03,gill05,mart06}. 
Recent results from the COSMOS survey are not able to provide statistically significant constraints on a possible enhanced activity associated to large scale structure \citet{gill08}.
 Still, X--ray data are relevant for
this kind of study, thanks to their high efficiency in identifying AGN.
In fact, before the current generation of X--ray observatories, like
\chandra, characterized by an high spatial resolution, the
investigation of the AGN distribution in and around clusters was based
only on the optical identification of AGN characteristic emission
lines in the galaxy spectra \citep{dres83,huch92,dres99}.  This
approach misses a large fraction of optically obscured AGN
\citep{marti02,marti06}, which, on the other hand, can be easily
identified by their hard X--ray emission (at least for intrinsic
column densities $N_H < 10^{24}$ cm$^{-2}$).

In the last years several studies were performed computing the
distribution of X--ray point sources in nearby galaxy clusters.  Significant
evidence was found of an excess of point sources covering a wide range of
redshift and cluster luminosity, around X--ray selected clusters
\citep{henr91,capp01,moln02,sun02,capp05}.  Here we perform a similar
study for the fields of the optically selected RCS clusters.

\subsection{Point Source Identification}

\begin{figure*}
\centering
\includegraphics[width=8.0 cm, angle=0]{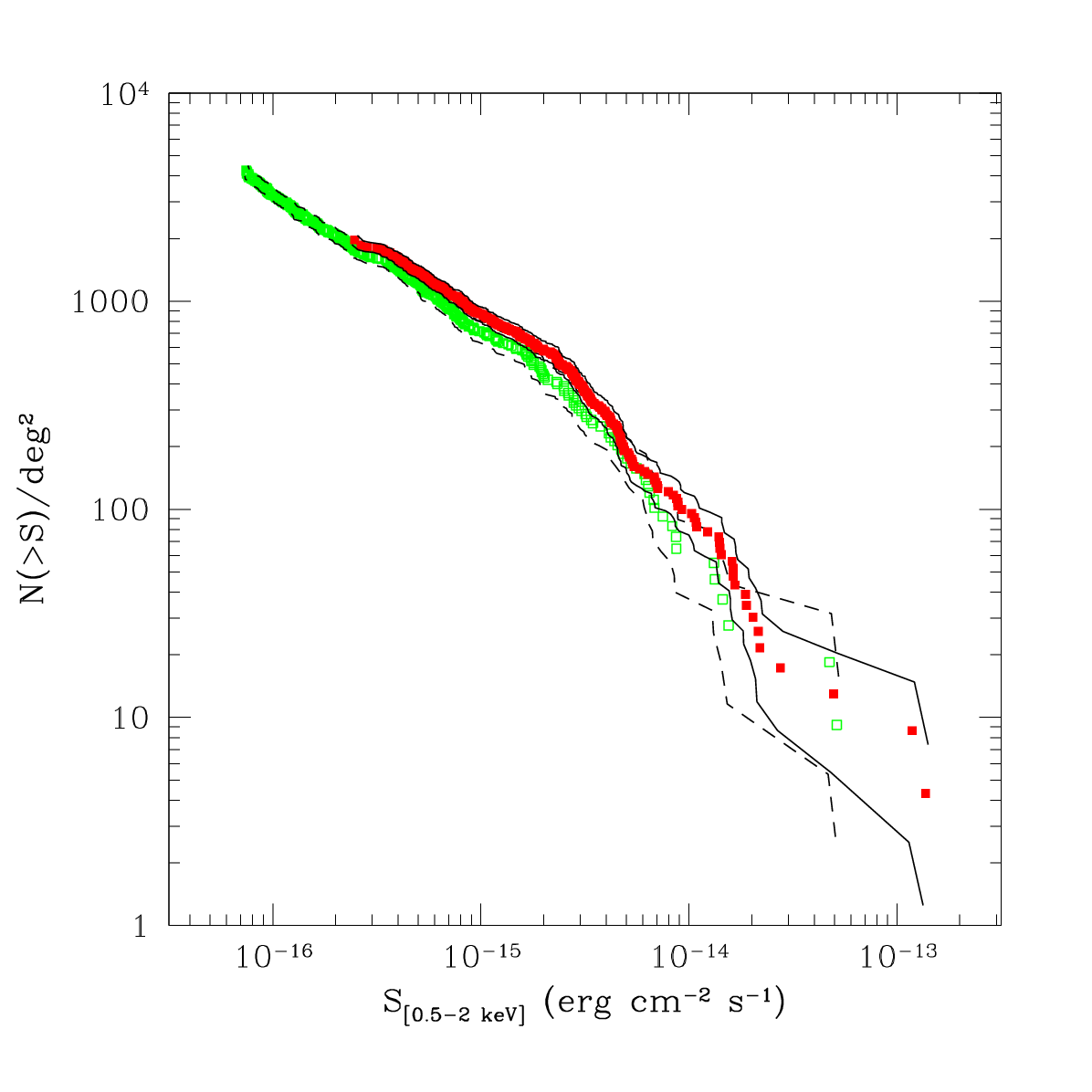}
\includegraphics[width=8.0 cm, angle=0]{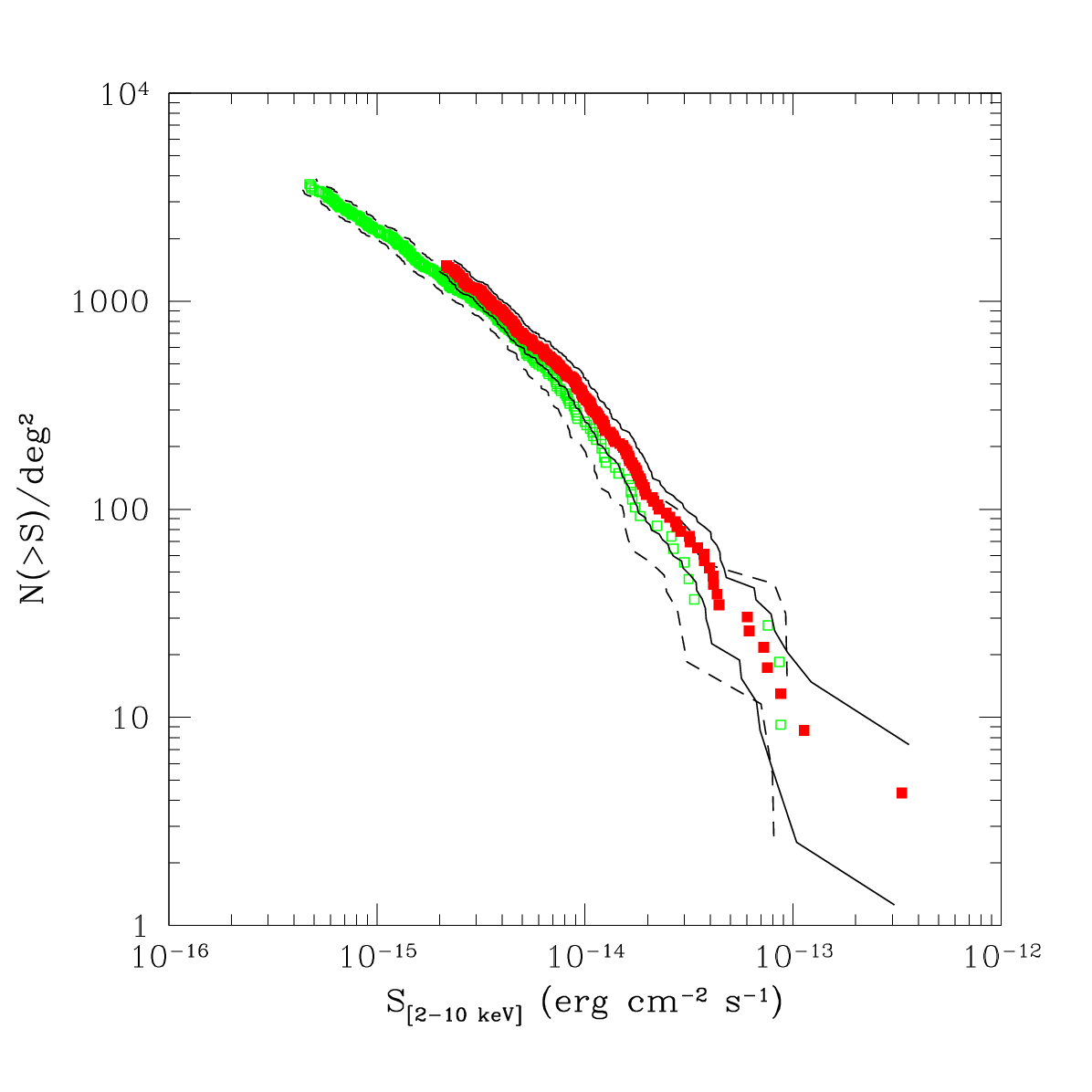}
\caption{LogN-LogS in the soft band for the observations with the ACIS-S detector.
The CDFS data is shown in green and the RCS data is shown in red. The solid (dashed) black line shows the error, 1$\sigma$,
of the RCS (CDFS) distribution.}
\label{fig:lognlogsaciss}
\includegraphics[width=8.0 cm, angle=0]{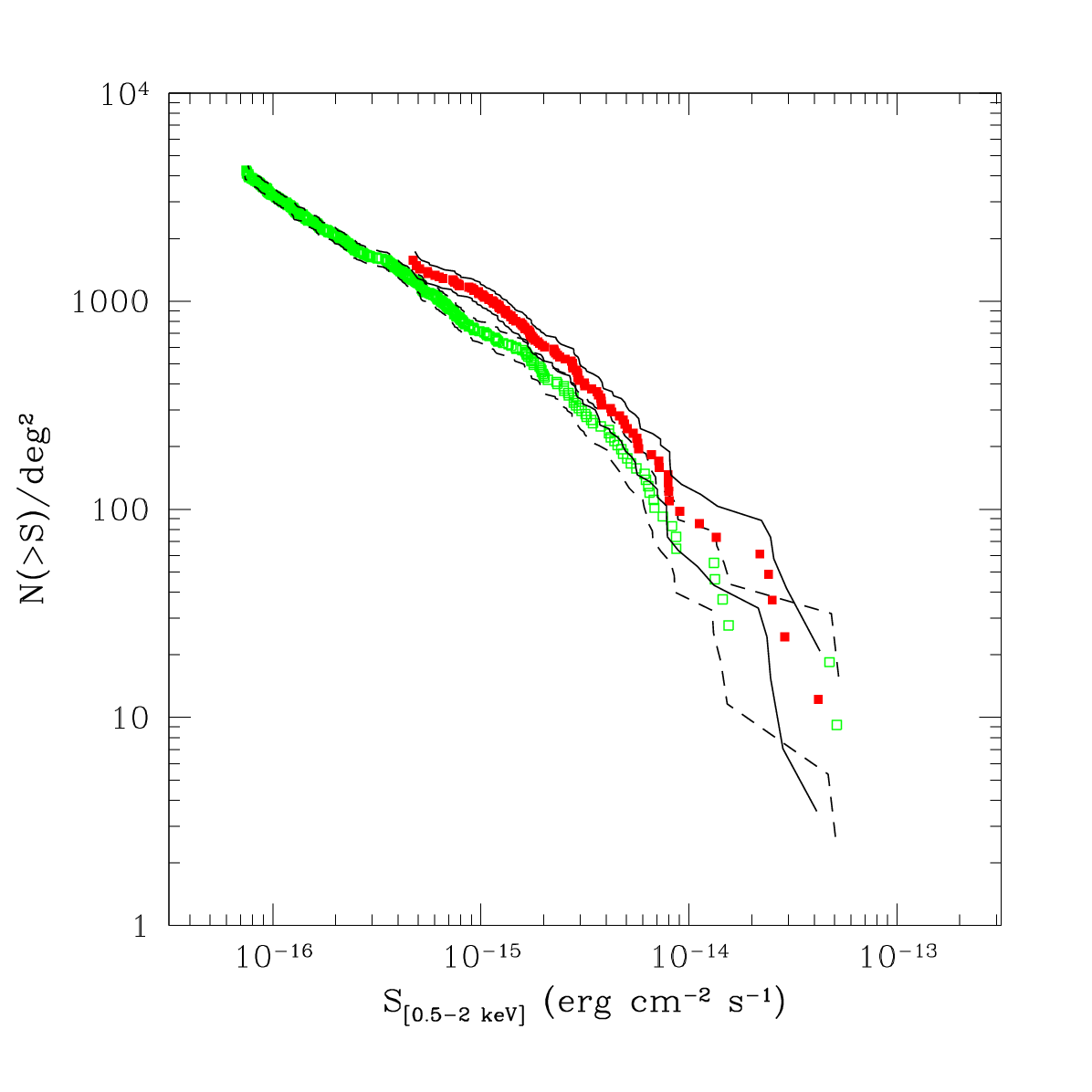}
\includegraphics[width=8.0 cm, angle=0]{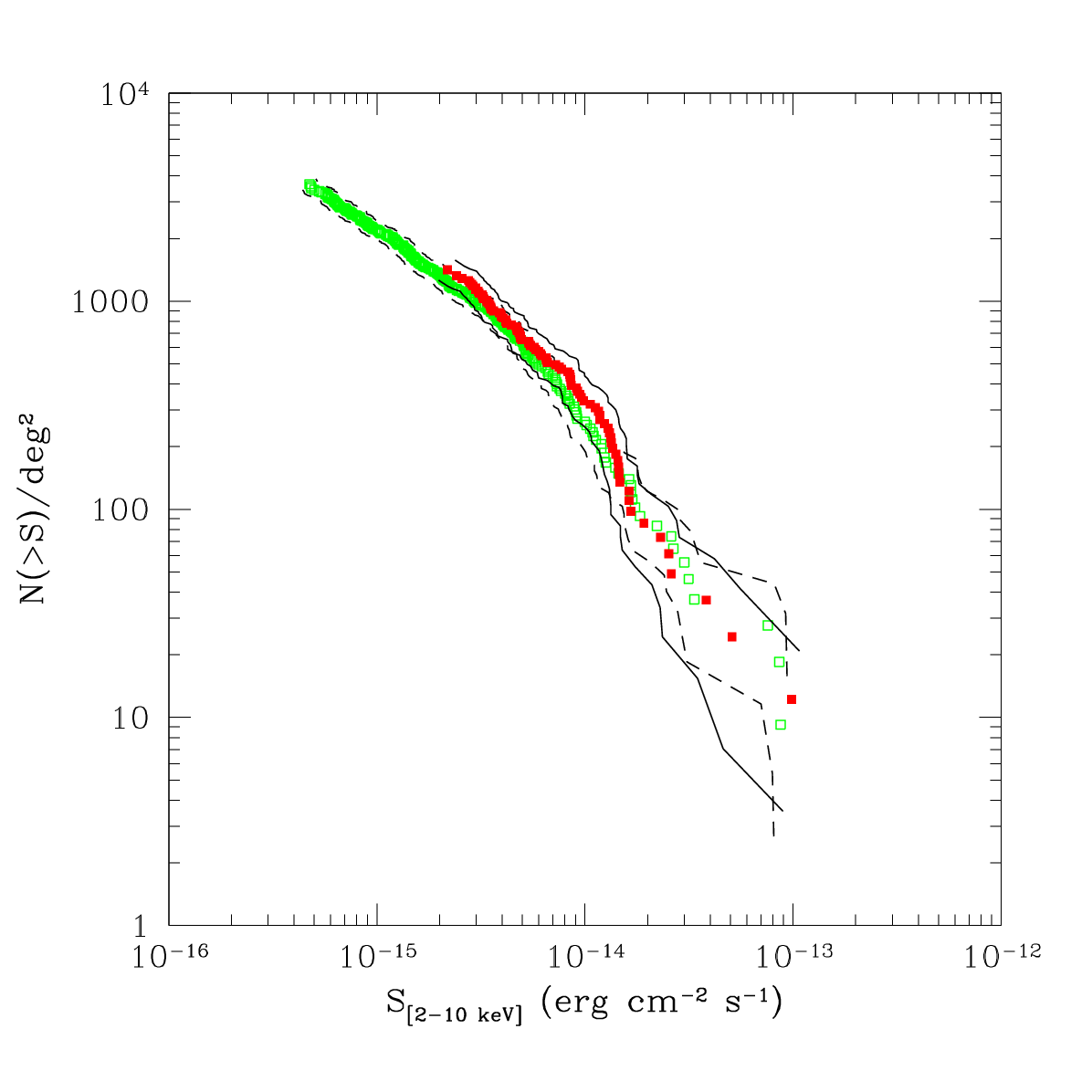}
\caption{LogN-LogS in the soft band for the observation with the ACIS-I detector.
The CDFS data is shown in green and the RCS data is shown in red. The solid (dashed) black line shows the error, 1$\sigma$,
of the RCS (CDFS) distribution.}
\label{fig:lognlogsacisi}
\end{figure*}

We used two different algorithms to identify point sources in RCS
fields. First we run {\tt wavdetect} implemented in CIAO
\citep{free02}, and then a modified version of the algorithm {\sl
Source--Extractor} ({\sl SExtractor}) \citep{bert96}.  In the second
case we performed a background subtraction using a background map
obtained by the same X--ray image where any point source candidate has
been previously removed.  We run both algorithms on the images
obtained in the soft (0.5--2.0 keV), hard (2.0--7.0 keV) and total
(0.5--7.0 keV) bands.

Finally, we combined in a single catalogue all the point sources
detected with $S/N > 2.1$ as measured from aperture photometry.  Here
we followed the same procedure used for \cdfs\
sources as described in \citet{giac01}. We measured the signal--to--noise
ratio of all the detected sources in the area of extraction of each
source, which is defined as a circle of radius $R_s=2.4\times FWHM$
(with a minimum of 5 pixels of radius).  The $FWHM$ was modeled as a
function of the off--axis angle to reproduce the broadening of the
PSF.  In each band a detected source has a $S/N \equiv S/\sqrt{S+2 B}>
2.1$ within the extraction area of the image.  Here $S$ is the source
net counts and $B$ is the
background counts found in an annulus with outer radius $R_s+12''$ and
an inner radius of $R_s+2''$, after masking out other sources,
rescaled to the extraction region.  Source counts were measured with
simple aperture photometry within $R_s$ in the soft and hard bands
separately.  Simulations have shown that such aperture photometry
leads to an underestimate of the source count rate by approximately
$4$\% \citep[see][]{tozz01}. We corrected such photometric bias
before converting count--rates in energy flux.

We removed by means of visual inspection double detections and
spurious detections due to occurrences that cannot be handled by the
detection algorithm (e.g.\ sources on the edges of the image or in
high background regions). Only in few cases we add by hand obvious sources
that were not identified by the algorithm (e.g.\ point sources missed because too close
to a bright source).  The final number counts distribution is
practically unaffected by these corrections.

We computed the effective sky--coverage at a given flux, which is
defined as the area on the sky where a source with a given net count
rate can be detected.  The
computation includes the effect of exposure, vignetting and
point spread function variation across the field of view.  The
count--rate to flux conversion factors in the 0.5--2.0 keV and in the
2.0--10.0 keV bands are computed using the response matrices.  We quoted
the fluxes in the canonical 2.0--10.0 keV band, as extrapolated from
counts measured from the 2.0--7.0 keV band.  The conversion factors are
computed for $\Gamma = 1.4$ at the aimpoint of each field, after
including the effect of the Galactic absorbing column (see Table
\ref{tab:rcslist}).

Before computing the energy flux of each source, the count rates are
corrected for vignetting and converted to the count rates
that would be measured if the source were in the aimpoint.  The
correction is simply given by the ratio of the value of the exposure
map at the aimpoint to the value of the exposure map at the source
position.  This is done separately for the soft and the hard band,
using the exposure maps computed at energies of 1.5 keV and 4.5 keV.
This procedure also accounts for the variations in exposure time
across the field of view.

In Table \ref{tab:rcssources} we show the solid angle covered by each
field, the number of sources detected with our criteria in the soft
and hard bands, the total number of sources in the combined catalogue
and the flux limits in the soft and hard bands.

\subsection{AGN Number Counts and Spatial Distribution in RCS Fields}

In order to investigate whether there is any excess of point sources
in and around RCS clusters with respect to the field, we computed the
point source number density as a function of flux ({\sl LogN--LogS})
and the spatial distribution of point sources.

The {\sl LogN--LogS} is defined as the logarithm of the number of
sources per unit of solid angle with flux greater than a given flux,
$Log(N(>S))$, as a function of the flux, {\sl S}.  The number density
$N(>S)$ can be computed as:
\begin{equation}
N(>S)=\sum_i \frac{1}{\omega(S_i)},
\end{equation}
where $\omega(S_i)$ is the value of the sky--coverage evaluated at the
flux of the source $S_i$ and the sum is over all the sources with flux
$S_i>S$.

We computed the total {\sl LogN--LogS} in the soft and hard bands
by summing all the fields, each one with its own sky--coverage (see
Figure \ref{fig:lognlogsaciss}).  In the same Figure we plot the data
of \cdfs\ (CDFS) from \citet{rosa02}. We
treated separately the field of RCS1417 (see Figure
\ref{fig:lognlogsacisi}), since it has been observed with
\hbox{ACIS--I} and therefore, having a much larger solid angle, has
many more point sources.

\begin{figure}
\centering
\includegraphics[width=8.0 cm, angle=0]{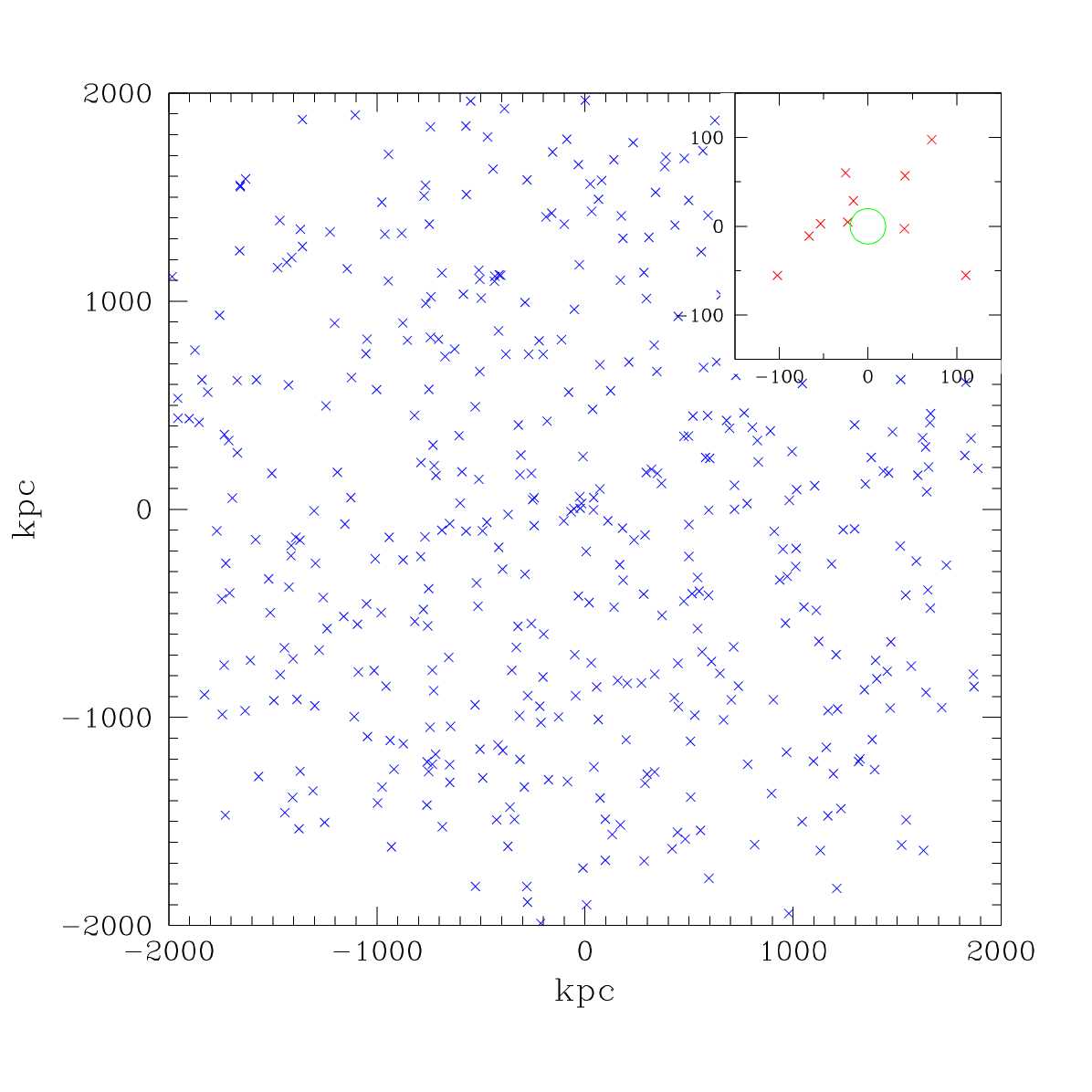}
\caption{Spatial distribution of point sources in the RCS fields.
This image was obtained by stacking the distribution of every single
field, matching the centroid of the X--ray emission of the clusters at the coordinates
(0,0). We stacked together only point sources with flux higher than the highest
flux limit. Distances are rescaled assuming the cluster redshift in each field. Every cross in the image is a point source.
The small panel in the top--right corner is a zoom of the inner region. The small green circle is centered in (0,0) and has a radius of 20 kpc.}
\label{fig:sourcesonly}
\end{figure}

\begin{figure}
\centering
\includegraphics[width=8.0 cm, angle=0]{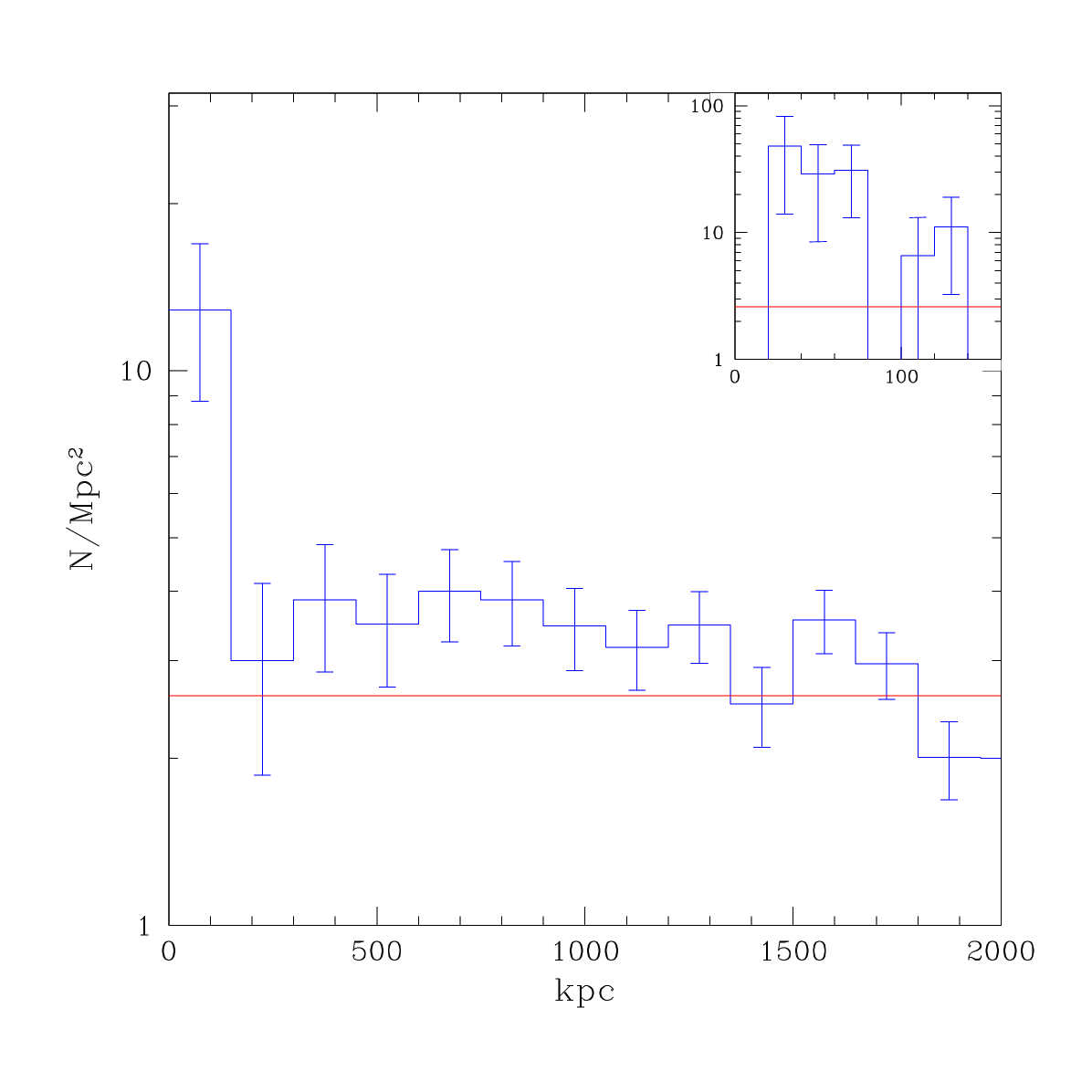}
\caption{Number density profile of point sources in the fields of the
RCS clusters. The profile was obtained from the superposition of the
spatial distribution of every single field joining in the clusters
position in the center of the image and binning the spatial
distribution by concentric rings. The horizontal line shows the mean
number density of point sources \textcolor{black}{obtained by the ratio between the total number of point sources in all fields and the total area of all fields}.  Error bars are at 1$\sigma$ level.
The small panel in the top--right corner is a zoom of the inner region re--binned with a bin size of 20 kpc.}
\label{fig:sourcesprofile2}
\end{figure}

Both in Figure \ref{fig:lognlogsaciss} and \ref{fig:lognlogsacisi} we
find a small excess of point sources in RCS fields at all fluxes.  In
the ACIS--S fields the surface density of point sources exceeds by
$\sim$20\% the value expected from the CDFS both in the soft and hard;
in the ACIS--I field by $\sim 40$\% in the soft and $\sim 15$\% in
the hard.  The significance of this excess is evaluated separately
after fitting the differential number counts with a single slope power
law.  We found the excess in the normalization to be significant
at 2.0 $\sigma$ and at 1.7 $\sigma$ confidence level in the soft and
hard band respectively for the ACIS--S fields, and at 2.0 $\sigma$
and at 0.9 $\sigma$ in the soft and hard band respectively for the
ACIS--I field.

This result is similar to what has been found by other authors.
\citet{capp01} found that the 0.5--2.0 keV source surface density (at a
flux limit of $1.5\times10^{-15}$ erg cm$^{-2}$s$^{-1}$) measured in
the area surrounding two clusters at $z \sim 0.5$ exceeds by a factor
of $\sim$ 2 the value expected in the field {\sl LogN--LogS}, with a
significance of $\sim 2 \sigma$.  Also \citet{capp05} found a factor of
$\sim 2$ over--density with a significance $ > 2 \sigma$ in 4 cluster
fields studying a sample of 10 high $z$ ($0.24 < z < 1.2$) clusters.
In a similar way \citet{bran07a} found a $\sim 2 \sigma$ excess of
sources in the cluster region at the bright end of the {\sl
LogN--LogS} for a sample of 18 distant galaxy clusters.  In summary,
there is growing evidence that there is an excess of X--ray sources
around clusters.  The low significance of this evidence is due to the
fact that the {\sl LogN--LogS} is dominated by all the sources along
the line of sight in the solid angle of each field (for example, at
$z\sim 1$ the 8 arcmin size of an ACIS--S field corresponds to 4 Mpc).
This result is expected, since the presence of a cluster implies
the presence of large scale structure with the consequent excess of
galaxies with respect to the field.  In order to investigate the
genuine enhancement of AGN activity around clusters, we would need to
perform an extensive spectroscopic follow up of the identified AGN to
select those associated with the cluster and compare their relative
density with respect to the field galaxies.  This would require an extensive survey of the galaxy population in the cluster and in the field, which is beyond the scope of this paper.

Without the spectroscopic information, we can use the spatial
distribution of the X--ray sources in order to isolate the AGN
actually associated with the cluster.  Therefore, we also computed the
number density of point sources in RCS fields as a function of the
distance from the center of the clusters.  Figure
\ref{fig:sourcesonly} shows the spatial distribution of point sources,
obtained by stacking the RCS fields overlapping \textcolor{black}{the centroid
of the X--ray emission of the clusters}. We computed the distances of the point sources from the
center of the cluster as if their redshifts were that of the cluster.
Since the flux limits slightly vary from one field to another (see
Table \ref{tab:rcssources}), we stack together only point sources with
flux higher than the highest flux limit, which is $3.6 \times 10^{-16}$
erg s$^{-1}$ cm$^{-2}$ in the soft band and $2.8 \times 10^{-15}$ erg
s$^{-1}$ cm$^{-2}$ in the hard band.

Altogether there are 12 point sources within 150 kpc ($\sim20$ arcsec)
from the center of the RCS clusters. Note that 150 kpc is smaller than the
extraction radii used for the cluster spectral analysis.
Almost every cluster has at least
one source within 150 kpc, RCS1107 and RCS2318 have two sources, while
only RCS2112 has none. We checked these 12 point sources one by one,
to make sure they are not spurious detections due to Poissonian
fluctuations in the thermal bremsstrahlung emission of the cluster
itself.

When the point source candidate are not clearly resolved we use, as a secondary criterion, the hardness ratio of its
emission.  Indeed, AGN emission is significantly harder than the
thermal bremsstrahlung emission from the ICM.  The hardness ratios of
these 12 point sources are listed in Table \ref{tab:rcssources2}. 
After comparing the hardness ratio of the point sources and the clusters, we removed two sources from the central bin: the
first one from RCS1419 and the second one from RCS1107.  In Figure
\ref{fig:sourcesprofile2} we plot the number of point sources as a
function of the distance from the center, computed at the redshift of
the cluster.  The excess with respect the mean density in the field is about a factor of $\sim6$ and it is significant
at $3\sigma$ confidence level  in the central bin (corresponding to the
innermost 150 kpc).

Both in Figure \ref{fig:sourcesonly} and \ref{fig:sourcesprofile2}
we performed a zoom of the inner region, to look more in detail the position of the AGNs with respect the center of the X--ray emission.
From the zoomed histogram in top--right corner of Figure \ref{fig:sourcesprofile2}
we found that no point sources are in the innermost 20 kpc.
Unfortunately, given the low number of point sources we have and the low statistic of our small sample
it looks difficult to make strong statements.
In conclusion, it is clearly evident an excess of point sources towards the center of the X--ray emission of the cluster, which seems not to affect the innermost (20 kpc) central region. On average the nearest point source
to the center of the cluster is distant about 70 kpc.
\correzione{Also \citet{hick07} with an independent analysis of RCS0224-0002 found a significant excess of point sources within $R_{200}$.}

A similar analysis was performed by \citet{rude05} with a sample of 51
clusters in the MACS survey. Their source list for all fields is
complete to a flux limit of $1.25 \times 10^{-14}$ erg s$^{-1}$
cm$^{-2}$ and in the radial profile of their source surface density
they found an evident excess in the central 0.5 Mpc by a factor of $\sim5$ significant at the
$8.0\sigma$ confidence level.  An analogous result was found by
\citet{bran07a} for a sample of 18 distant galaxy clusters.

\begin{figure*}[htp]
 	\centering
 	\includegraphics[width=8.0 cm, angle=0]{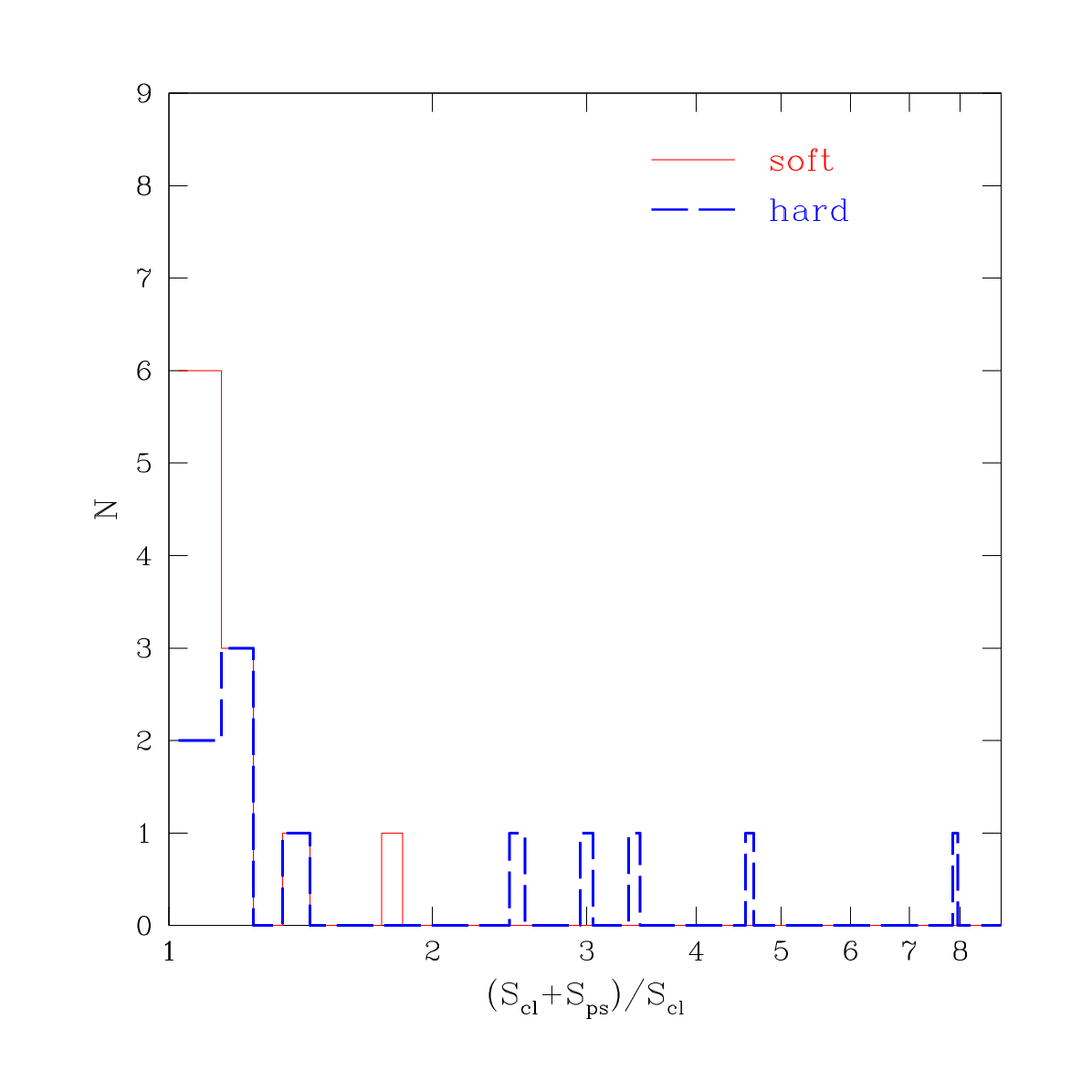}
	\includegraphics[width=8.0 cm, angle=0]{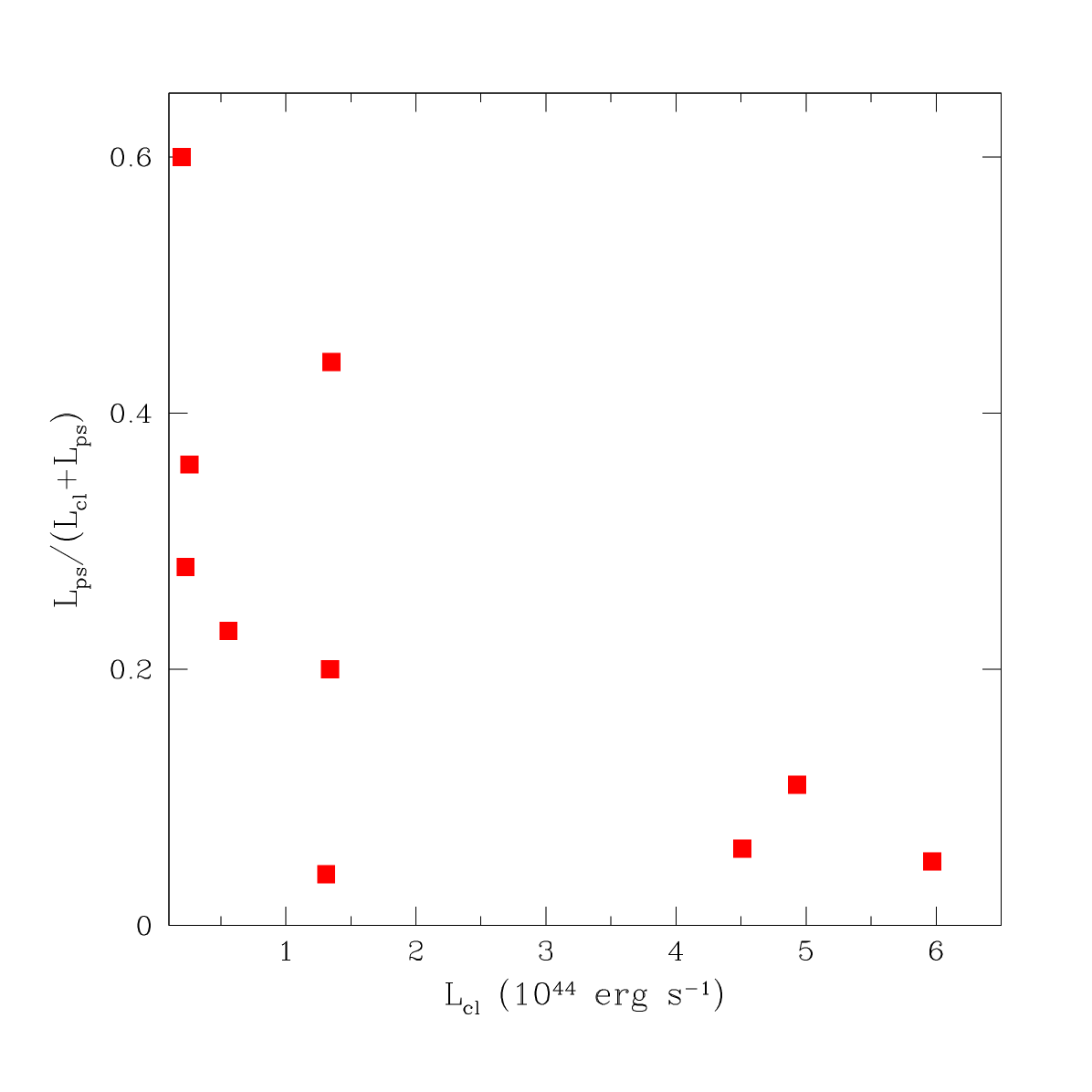}
	\caption{Left panel: histogram of the ratios between the flux
	of the cluster plus the flux of the point sources embedded and
	the flux of the extended cluster emission.  Number larger than
	2 indicates that the emission is dominated by AGN rather than
	the ICM.  The solid red line is for the soft band and the
	dashed bold blue line is for the hard band.  \textcolor{black}{Right panel: ratio
	between total 0.5-10.0 keV luminosities of point sources and ICM
	luminosity plus point sources luminosity as a function of the ICM
	luminosity itself.  Lower luminosity clusters tend to be more
	contaminated by point--source emission.}}
 	\label{fig:Lcl_Lps}
\end{figure*}

\textcolor{black}{
In the last columns of the Table \ref{tab:rcssources2}, we show for
all the point sources of the central bin (inner 150 kpc) the luminosities that the sources would have
at the cluster redshift. In Figure \ref{fig:Lcl_Lps} we show the
comparison between the total emissions within the extraction radius (ICM plus point sources) and the
ICM emission.  In the left panel we investigated the contribution of
the point sources to the total flux in the soft and hard energy bands
separately, by the ratio $(S_{cl}+S_{ps})/S_{cl}$, where $S_{cl}$ and
$S_{ps}$ are the flux of the cluster and of the point sources
respectively. 
}

\textcolor{black}{
As expected the point sources contribution to cluster flux and luminosity is more
prominent in the hard band (as one would naturally expect from the
power--law AGN spectrum harder than a typical thermal spectrum); instead in the soft band for the great majority of cases the point sources contribution to total flux is less than 20\%.
}

\textcolor{black}{
In the right panel we plotted the quantity
$L_{ps}$/($L_{cl}+L_{ps}$) \emph{versus} $L_{cl}$, where $L_{cl}$ and
$L_{ps}$ are the luminosity in the 0.5--2.0 keV band of the cluster and of the point sources
respectively. Large values of this ratio
indicate that the emission is dominated by the AGN.
The quantity on the {\sl y} axis can be interpreted as percentage contamination
of cluster luminosity due to point sources.
As one can see the point sources contamination in greater for clusters with lower luminosity. On the other hand
the total luminosity is always dominated by
the ICM except in one case.
}

These considerations are similar to what was obtained by
\citet{bran07b}, who conclude that point sources located
within the ICM region may affect considerably the estimates of X--ray
observable showing that the point source contribution must be
removed.  However, contamination in the soft band is not severe, and
we do not find indication of a possible population of clusters which
would have been missed in X--ray due to the presence of bright AGN.

\section{Conclusions}

We studied the thermodynamical and chemical X--ray properties of the
ICM of a galaxy cluster sample of the Red--Sequence Cluster Survey, in
the redshift range \hbox{$0.6<z<1.2$}. We detected emission for the
majority of the clusters, except for three, for which we have only
marginal detection at $\sim3\sigma$.  In general  we found that the slope
of the $\lxtx$ relation of the RCS clusters is in agreement with
that of X--ray selected clusters, while the normalization is a factor of 2 lower
at high confidence level, in agreement with the result of \citet{hick08} in their independent analysis. Only the three marginally detected RCS clusters
seem to have lower luminosities with respect to the RCS $\lxtx$, but only if they have $kT>3$ keV.
Unfortunately, for this sample the
statistic is too poor to draw any conclusion about possible evolution
of the $L_{\mathrm X}-T_{\mathrm X}$ relation with redshift.
Particularly our data are fitted with $L_{\mathrm X}/T_{\mathrm X}^{\alpha} \propto (1+z)^{0.2 \pm 0.2}$, consistent with no evolution.

Concerning the Fe abundance in RCS clusters we found that also for
this sample of optically selected clusters, the ICM was already
enriched with metals at a level comparable with X--ray selected
clusters at high redshift \citep{bale07}.

Thanks to the high spatial resolution of the \chandra\ satellite,
we also investigate the point source distribution near the region of
diffuse cluster emission for RCS with respect to the field.  The number
counts of point sources as a function of the flux
show a significant excess at all fluxes with respect to the data of the
CDFS.  We quantified this excess between 15\% and 40\% with a
significance of $2\sigma$.  This result is in agreement with that
found by other authors \citep{capp01,bran07a}.

The spatial distribution of point sources in the field of RCS clusters
shows a factor of $\sim6$ over--density at $3\sigma$ confidence level in the inner
150 kpc.  The contribution of these point sources to the X--ray
emission is limited to few percent in the soft band, showing that the
contamination from AGN in the soft band is not severe at least for
cluster with $L_X > 10^{44}$ erg s$^{-1}$ up to $z\sim 1$.

\acknowledgements

We acknowledge the anonymous referee for substantial comment which helped to improve this paper.
We wish to thank E.\ Ellingson for discussion and A.\ Hicks for helpful comments.
We acknowledge financial contribution from contract ASI--INAF I/023/05/0 and ASI-INAF I/088/06/0.
PT and SB acknowledge financial contribution from the PD51 INFN grant.

\clearpage

\appendix

\section{Properties of Individual Clusters}

\subsection*{RCS1419+5326}

\begin{figure*}
\centering
\includegraphics[width=0.35\textwidth, angle=270]{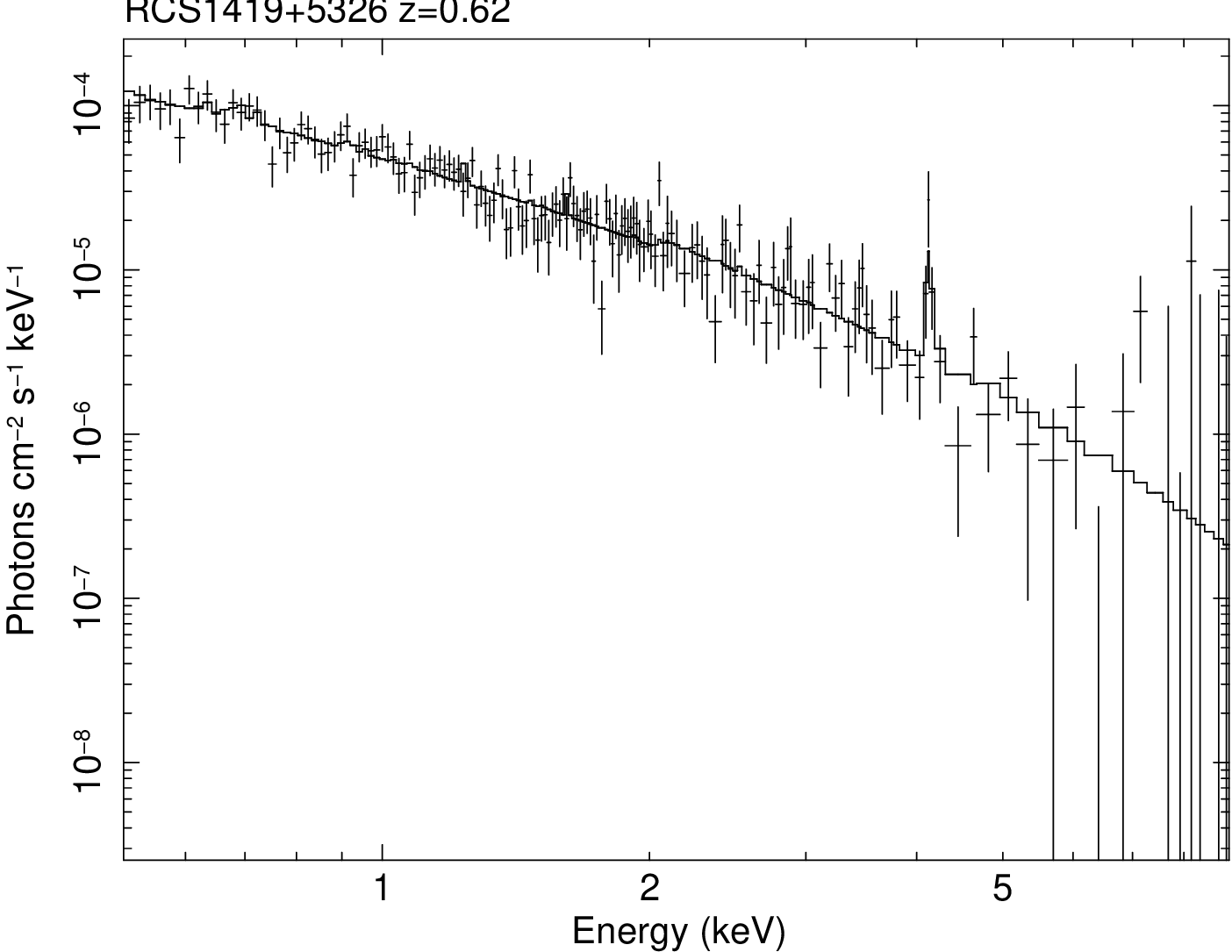}
\includegraphics[width=0.35\textwidth, angle=270]{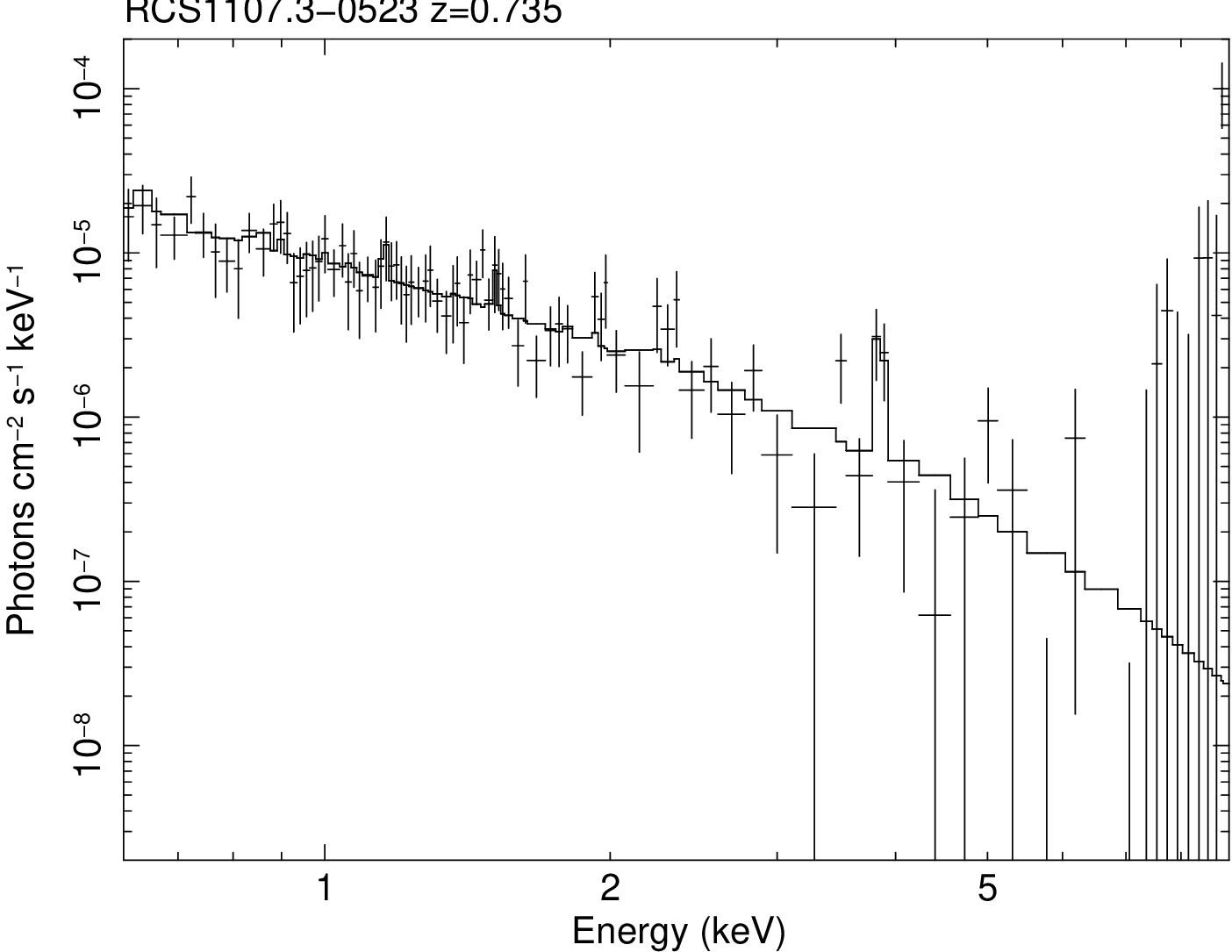}
\includegraphics[width=0.35\textwidth, angle=270]{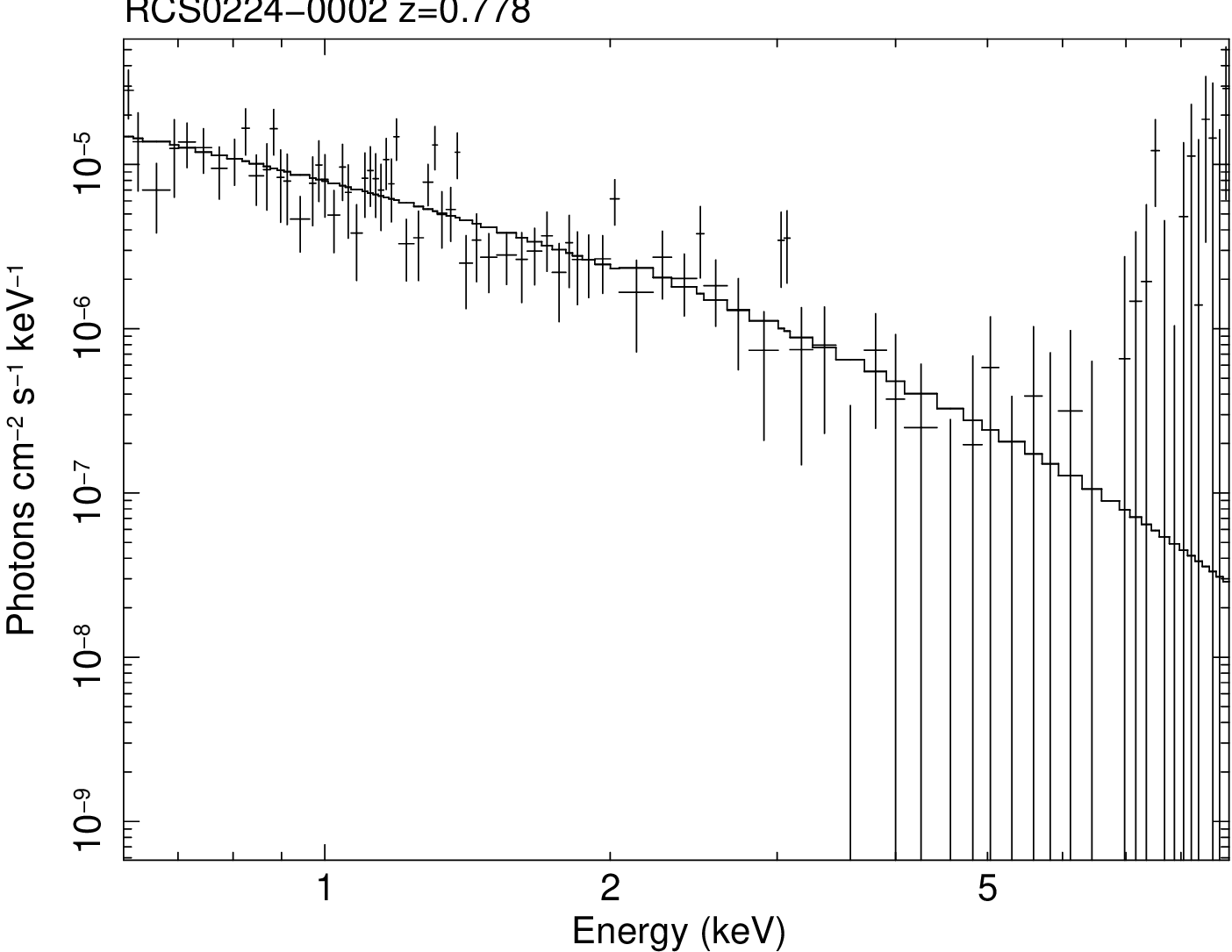}
\includegraphics[width=0.35\textwidth, angle=270]{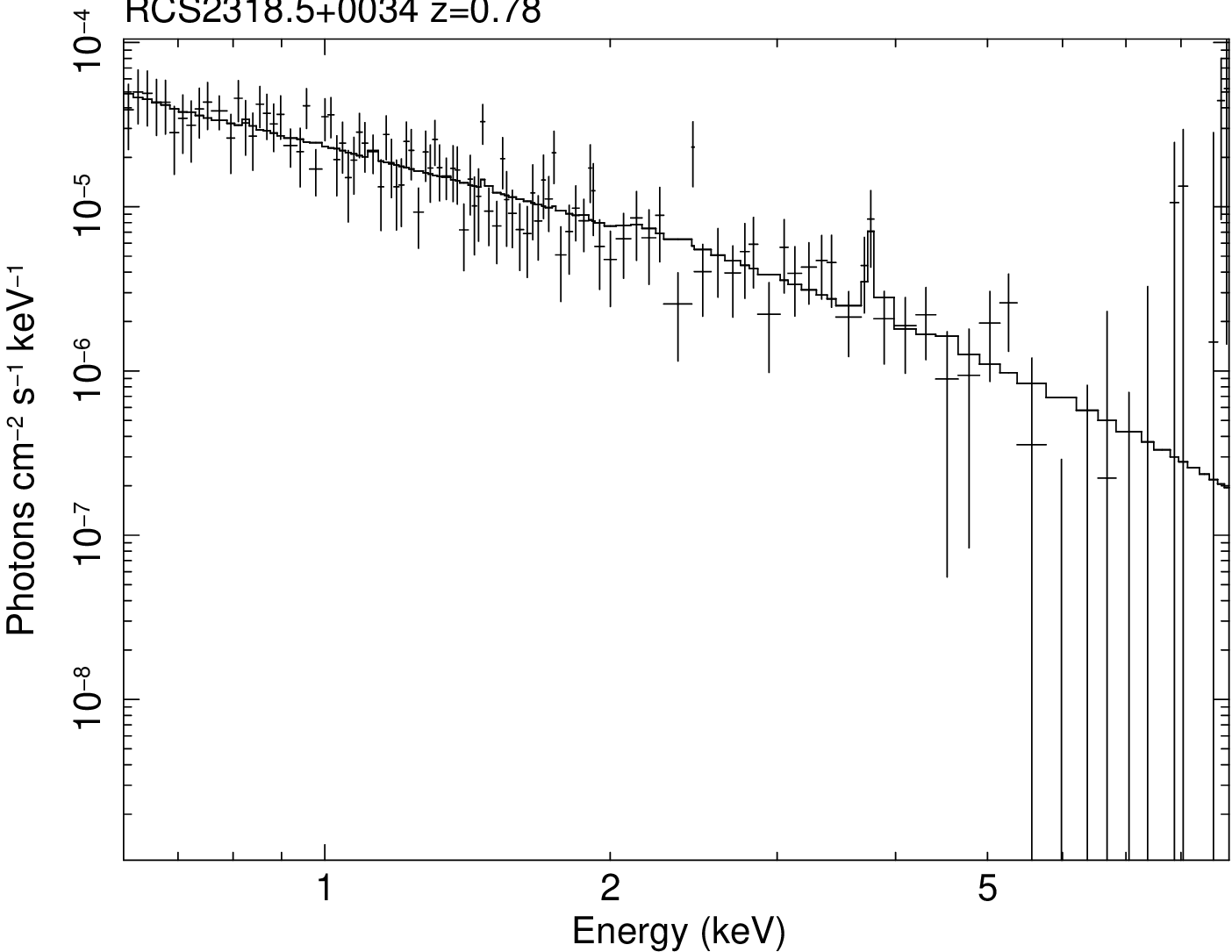}
\caption{Unfolded spectra of RCS1419+5326, RCS1107.3-0523, RCS0224-0002 and RCS2318.5+0034. The solid lines are the best fit of the model.}
\label{fig:spectra1}
\end{figure*}

RCS1419+5326 is the lowest z RCS cluster ($z=0.62$) in our sample and
was observed with two pointings, ObsID 3240 with nominal exposure time
10 ks and ObsID 5886 with 50 ks. The extraction radius is 37$''$,
corresponding to 252 kpc.  RCS1419+5326 is detected with the highest
number of net counts, $2320 \pm 60$, corresponding to a
signal--to--noise ratio $\sim 38$. In the spectrum (Figure
\ref{fig:spectra1} top--left panel) there is a clear evidence of the
Iron $K_\alpha$ line at $\sim 4.14$ keV (observing frame).  The
resulting Fe abundance is $\XFe=0.29_{-0.11}^{+0.06} \XFeS$, with a
temperature $kT=5.0_{-0.4}^{+0.4}$ keV and bolometric luminosity
$L_{\mathrm X}=4.63\pm0.12\times10^{44}$ erg s$^{-1}$.  As pointed out
by \citet{sant08} RCS1419+5326 is a cool--core cluster. Therefore we
repeated the spectral analysis masking the cool--core roughly
corresponding to the inner 80 kpc. \textcolor{black}{In this case the best fit
temperature is $kT=5.2_{-0.5}^{+0.7}$ keV and the resulting Fe abundance is substantially unchanged
$\XFe=0.23_{-0.14}^{+0.12}$. We corrected the bolometric luminosity
for the removed core emission fitting the radial surface brightness profile with a $\beta$ model
\begin{equation}
I(r)=A \left[ 1+ \left(\frac{r}{r_c}\right)^2 \right]^{-3\beta+1/2}
\end{equation}
where $A$ is the amplitude at $r=0$ and $r_c$ is the core radius.
The best fit values are $r_c=17''$ and $\beta=0.78$.
Extrapolating the profile in the masked inner region,
the corrected bolometric luminosity is $L_{\mathrm
X}=3.71\pm0.15\times10^{44}$ erg s$^{-1}$.
}

\subsection*{RCS1107.3+0523}

This cluster at $z=0.735$ was observed with two pointings, ObsID 5825
and ObsID 5887, both with nominal exposure 50 ks. The extraction
radius is 28.4$''$, corresponding to 207 kpc.  The total number of net
counts is $710 \pm 40$.  The best fit temperature is
$kT=4.3_{-0.6}^{+0.5}$ keV resulting in a bolometric luminosity of
$L_{\mathrm X}=1.34\pm0.08\times10^{44}$ erg s$^{-1}$.  The Iron line is well visible at $\sim3.86$ keV (observing
frame).  This cluster shows the highest value of Fe abundance in our
sample, $\XFe=0.67_{-0.27}^{+0.35} \XFeS$

\subsection*{RCS1325+2858}

This cluster at $z=0.75$ has been observed with two pointings: ObsID
3291 with 31.5 ks and ObsID 4362 with 45 ks. The extraction radius is
29.7$''$, corresponding to 218 kpc.  We detected $90\pm 30$ net
counts.  As one can see from the spectrum (Figure \ref{fig:spectra2}
bottom--left panel) the signal is dominated by the background above 2 keV, so we found only
an upper limit to the Fe abundance of $\XFe=0.09_{-0.09}^{+0.66} \XFeS$.  The best fit
temperature is $kT=1.8_{-0.6}^{+1.2}$ keV and the resulting bolometric luminosity
is $L_{\mathrm X}=0.23\pm0.07\times10^{44}$ erg s$^{-1}$.

\subsection*{RCS0224-0002}

For this cluster at $z=0.778$ we analyzed two observations: ObsID 3181
with nominal exposure 15 ks and ObsID 4987 with 100 ks.  The
extraction radius is 36.7$''$, corresponding to 273 kpc.  The total
number of net counts is $740 \pm 50$.  The best fit temperature is
$kT=5.1_{-0.8}^{+1.3}$ keV and the bolometric luminosity is $L_{\mathrm
X}=1.31\pm0.10\times10^{44}$ erg s$^{-1}$.  In this case we
find only an upper limit to the Fe abundance $\XFe < 0.15 \XFeS$

\subsection*{RCS2318.5+0034}

RCS2318.5+0034 ($z=0.78$) was observed with a single observation,
ObsID 4938, with an exposure time of 50 ks.  The extraction radius is
40.6$''$, corresponding to 302 kpc. We detect $ 970 \pm 50$ net counts
for a signal--to--noise ratio of $\sim19$.  The Iron $K_\alpha$ line
at $\sim 3.76$ keV observing frame is very clear. The resulting Fe
abundance is $\XFe=0.35_{-0.22}^{+0.20} \XFeS$, with a best fit
temperature of $kT=7.3_{-1.0}^{+1.3}$ keV and a bolometric luminosity
$L_{\mathrm X}=4.51\pm0.23\times10^{44}$ erg s$^{-1}$.

\subsection*{RCS1620+2929}

For RCS1620+2929 at $z=0.87$ we have only one pointing, ObsID 3241,
with a nominal exposure time of 35 ks. We detected $ 190 \pm 20$ net
counts, extracted in a region of radius 29.5$''$, corresponding to 227
kpc.  The best fit temperature is $kT=4.6_{-1.1}^{+2.1}$ keV, the bolometric
luminosity is $L_{\mathrm X}=1.35\pm0.18\times10^{44}$ erg s$^{-1}$
and the Fe abundance is $\XFe=0.33_{-0.33}^{+0.60} \XFeS$.

\subsection*{RCS2319.9+0038}

RCS2319.9+0038 at $z=0.9$ was observed with four pointings: ObsID
5750, 7172, 7173 and 7174 with a nominal exposure time of
21 ks, 18 ks, 21 ks and 15 ks, respectively.  This is the most distant cluster in
our sample for which we have a high number of total net counts, $1490
\pm 60$, for a signal--to--noise ratio of $\sim 22$. It is also the
cluster with the largest extraction radius, 45.6$''$ arcsec (356 kpc),
the highest bolometric luminosity, $L_{\mathrm
X}=5.97\pm0.26\times10^{44}$ erg s$^{-1}$. The best fit temperature is
$kT=5.3_{-0.5}^{+0.7}$ keV.  The spectrum (Figure
\ref{fig:spectra2} bottom--right panel) show a clear $K_\alpha$ line
at $\sim3.53$ keV observing frame. The best fit Fe abundance
is $\XFe=0.60_{-0.18}^{+0.22} \XFeS$.

\subsection*{RCS0439.6-2905}

\begin{figure*}
\centering
\includegraphics[width=0.35\textwidth, angle=270]{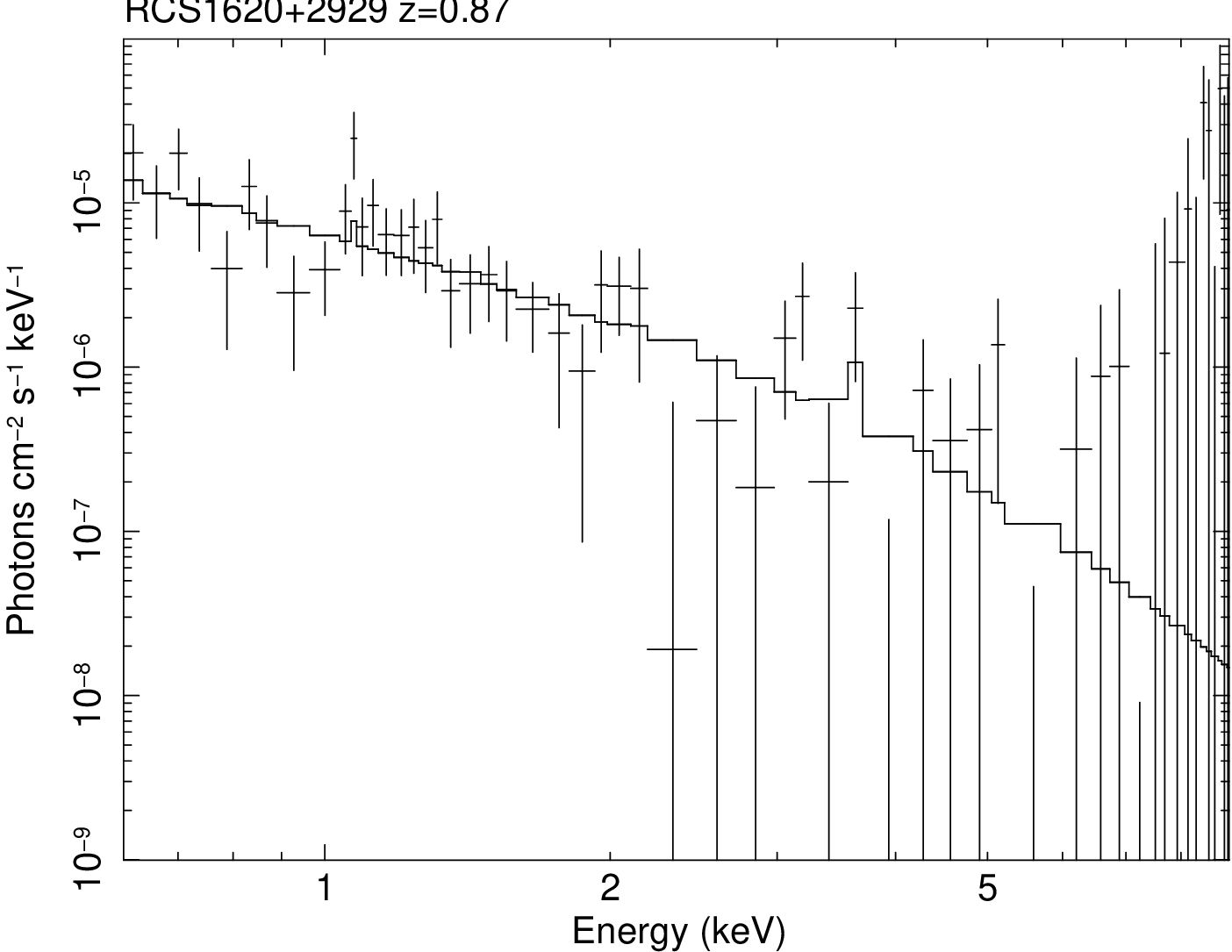}
\includegraphics[width=0.35\textwidth, angle=270]{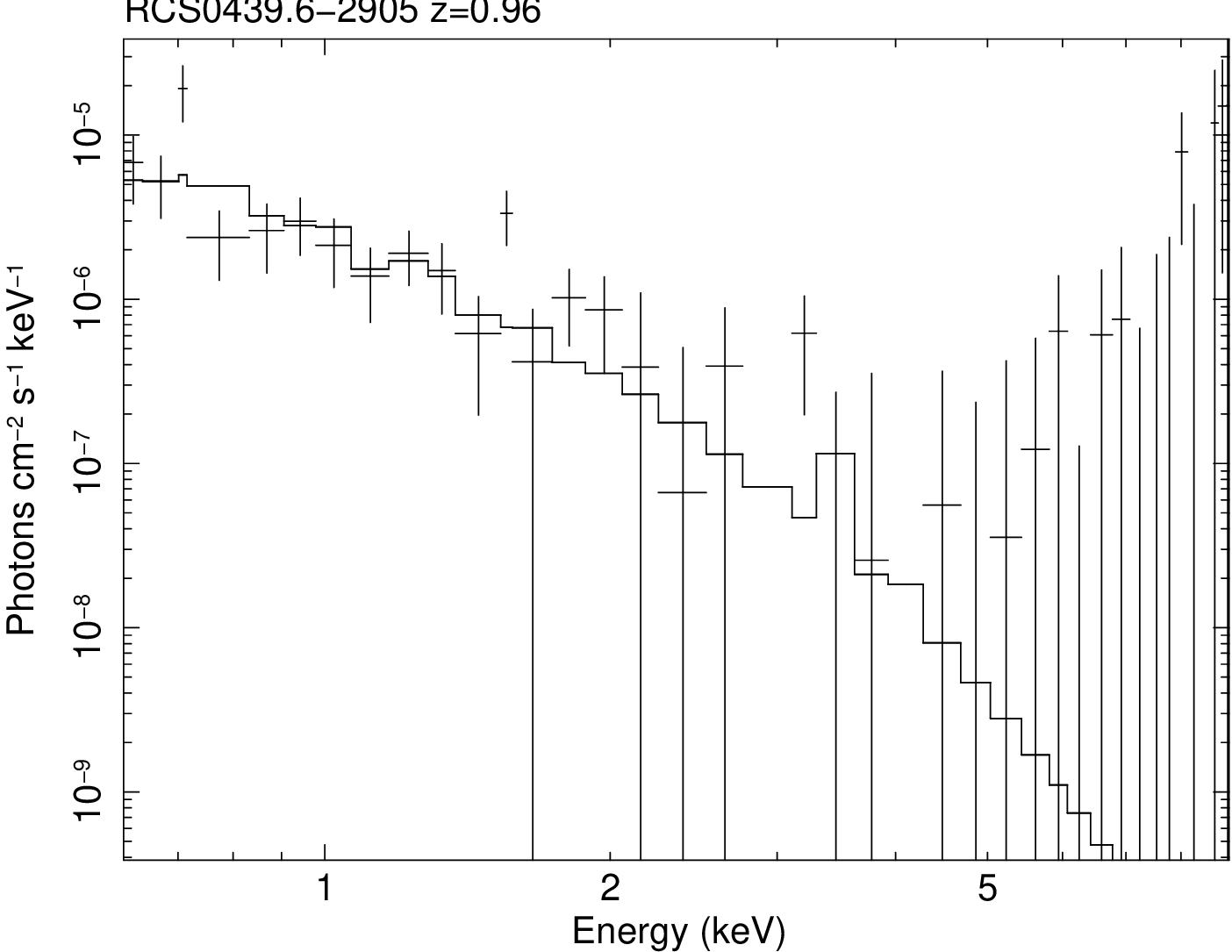}
\includegraphics[width=0.35\textwidth, angle=270]{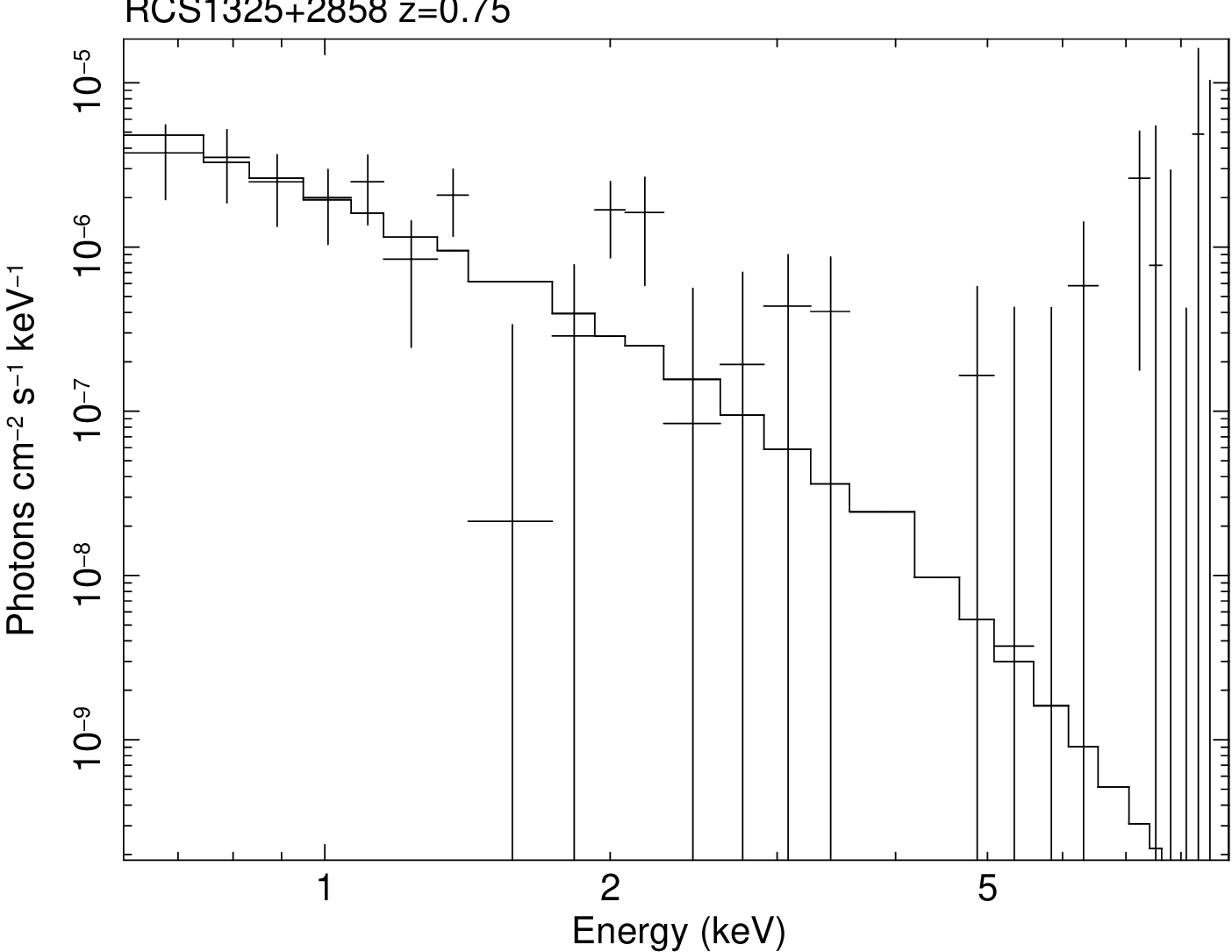}
\includegraphics[width=0.35\textwidth, angle=270]{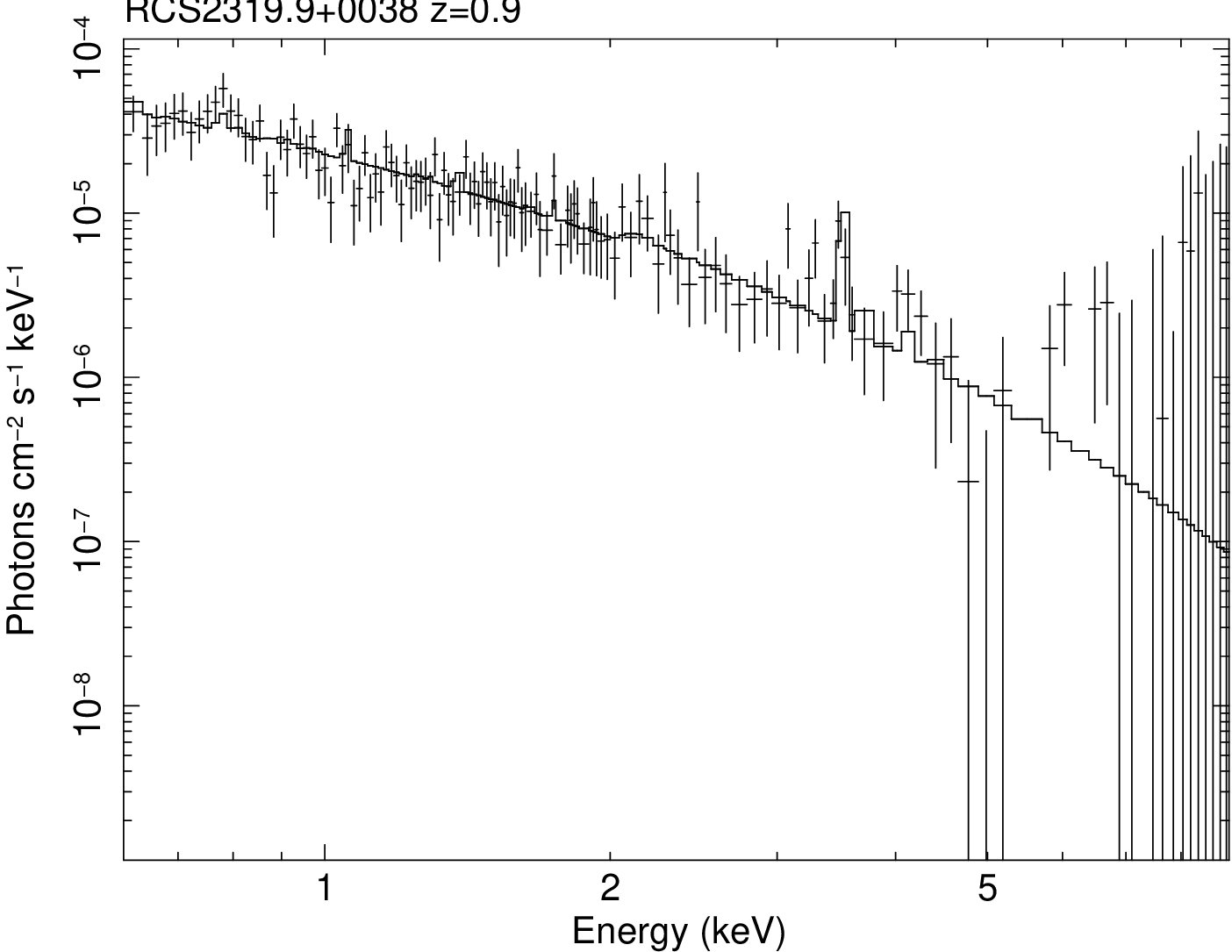}
\caption{Unfolded spectra of RCS1620+2929, RCS0439.6-2905, RCS1325+2858 and RCS2319.9+0038. The solid lines are the best fit of the model.}
\label{fig:spectra2}
\end{figure*}

RCS0439.6-2905 ($z=0.96$) is observed with two pointings: : ObsID 3577
with nominal exposure time of 85 ks and ObsID 4438 with 30 ks.  We
detected $220 \pm 30$ net counts.  The extraction radius is 24.7$''$,
corresponding to 196 kpc.  The best fit temperature is
$kT=1.8_{-0.3}^{+0.4}$ keV and the luminosity $L_{\mathrm
X}=0.56\pm0.09\times10^{44}$ erg s$^{-1}$. As one can see
from the spectrum (Figure \ref{fig:spectra2} top--right panel) the
signal is dominated by the background above 2 keV making impossible to
identify the Iron $K_\alpha$ line, so we found only an upper limit to
the Fe abundance of $\XFe=0.44\pm0.27 \XFeS$.

\subsection*{RCS1417+5305, RCS2112.3-6326, RCS2156.7-0448}

\begin{figure}
 \includegraphics[width=8.0 cm, angle=0]{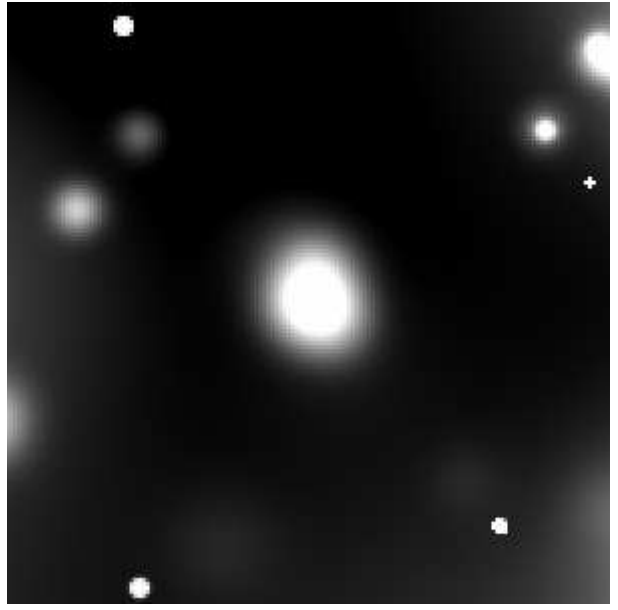}
	\caption{Soft merged image for the three undetected clusters RCS1417+5305, RCS2112.3-6326, RCS2156.7-0448. The size of the image is about $1'\times1'$.}
\label{fig:invisible}
\end{figure} 

In these three clusters, we are not able to detect the diffuse
emission in order to perform a standard spectral analysis. Therefore,
we selected a circular region with radius $\sim 20$ arcsec (20 pixel)
centered in the optical coordinates of each cluster.  In order to
perform a simplified spectral analysis, we constrained the temperature
to be in the range 1.0--8.0 keV and the metallicity at 0.3$\XFeS$,
thawing only the normalization. In this way we obtain a
reliable upper limit to the bolometric luminosity.

RCS1417+5305 ($z=0.968$) is the only RCS cluster in our sample observed
with ACIS-I (ObsID 3239) with an exposure time of 70 ks. The aperture
photometry in the extraction region, corresponding to 156 kpc, gives
$37 \pm 11$ net counts, consistent with zero within $3 \sigma$.
Freezing the temperature in the range 1.0--8.0 keV and the metallicity
to $0.3 \XFeS $, we obtain an upper limit to the bolometric
luminosity of $L_{\mathrm X}<0.29\times10^{44}$ erg s$^{-1}$.

RCS2112.3-6326 was observed with a single observation, ObsID 5885,
with nominal exposure time of 70 ks.  This is the most distant
cluster in our sample at $z=1.099$. The
extraction radius corresponds to 161 kpc. The total number of net
counts is $47 \pm 20$, implying a positive detection at $2 \sigma$ only.
The upper limit to the bolometric luminosity is $L_{\mathrm
X}<0.21\times10^{44}$ erg s$^{-1}$.

RCS2156.7-0448, whose redshift is $z=1.080$, was observed with two pointings: ObsID 5353 with 40 ks
and ObsID 5359 with 35 ks.   The aperture photometry in the extraction region,
corresponding to 160 kpc, gives $60 \pm 20$ net counts, consistent with
zero within $3\sigma$.  The upper limit to the bolometric luminosity is
$L_{\mathrm X}<0.22\times10^{44}$ erg s$^{-1}$.

In order to better constrain the average diffuse emission of these
three clusters, we merged together the three images
overlapping the optical centers of the clusters.  From the merged
image we chose the same extraction radius as before ($\sim 20$ arcsec)
and we performed the standard spectral analysis, fitting the
background subtracted spectrum with three free parameters, with the aim of obtaining a
mean value for the thermodynamical properties of these three clusters.
Unfortunately, even in this case the fit does not converge.  However,
from aperture photometry we measure \hbox{$274 \pm 29 $} total net
counts, implying a clear detection of diffuse emission from the hot ICM.
 Figure \ref{fig:invisible} is the soft
merged image, where the diffuse emission contributed by the sum of the
three clusters is clearly visible in the center.

Since the fit does not converge, to evaluate also a mean bolometric luminosity for these three clusters
from the total net counts, we assume the typical thermal spectrum for each cluster with the same redshift and exposure map,
in order to have the conversion factors from net counts to bolometric luminosity.
The temperature is fixed at 4 keV and the metallicity at 0.3 $\XFeS$.
Then we computed an effective conversion factor from these three single conversion factors, weighing them
with the net counts of the individual clusters. Multiplying 274 net counts by the effective conversion factor,
we found $\langle L_{\mathrm X}\rangle=0.2\pm0.06\times10^{44}$ erg s$^{-1}$ as the mean bolometric luminosity for one cluster
in agreement with upper limits found
from the single cluster analysis.

\end{document}